\documentclass[prb,twocolumn,aps,superscriptaddress,longbibliography,amsmath,amssymb]{revtex4-2}

\usepackage{graphicx}
\usepackage{braket}
\usepackage{hyperref}
\hypersetup{hidelinks}
\allowdisplaybreaks[2]
\begin{document}

\title{Eigenstate-selective entangled two-photon absorption
in Monolayer WSe$_2$}

\author{Minseok A. Jang}
\author{Hongki Yoo}
\email{h.yoo@kaist.ac.kr}
\affiliation{Department of Mechanical Engineering, KAIST, Daejeon, Korea}


\begin{abstract}
	We show that the Bell-state phase of a polarization-entangled photon pair controls the biexciton eigenstate distribution produced by entangled two-photon absorption in monolayer WSe$_2$. In a frequency-nondegenerate ladder scheme, two independent valley pathways ($K$ and $K'$) share no intermediate state, so the biphoton phase sets the relative amplitude between them. Within the valley-symmetric limit, the phase dependence is contained in the prefactors $1\pm e^{i\varphi}$, which are independent of the material response. The resulting selection rule partitions the excitation among biexciton eigenstates according to the Bell-state phase $\varphi$. The symmetric Bell state ($\varphi = 0$) selectively drives bright eigenstates, while the antisymmetric state ($\varphi = \pi$) drives the exchange-dark eigenstate. No separable polarization state reproduces this $\varphi$-dependent eigenstate distribution. At equal total pair flux, the dark-state pumping rate of the Bell state is twice the upper bound for any separable polarization state. Exceeding this bound certifies polarization entanglement of the source. Including valley dephasing, intervalley scattering, source imperfections, and residual amplitude errors at 4~K, the estimated phase-scan visibility of the eigenstate-resolved pair-absorption rates is $V\gtrsim0.95$ for broadband spontaneous parametric down-conversion (entanglement time $T_e \sim 100$~fs) with high source concurrence and picosecond-scale valley coherence.
\end{abstract}

\maketitle

\section{\label{sec:intro}Introduction}

Entangled two-photon absorption (ETPA) refers to a two-photon absorption process driven by an entangled photon pair, typically generated by spontaneous parametric down-conversion (SPDC). Because the photons are entangled in time-energy, polarization, or both, ETPA is expected to differ qualitatively from classical two-photon absorption. The entangled signal scales linearly rather than quadratically with photon flux at low intensities, and the entanglement degrees of freedom can in principle be used to control the excitation pathway~\cite{dorfman2016,schlawin2018,raymer2021,gu2020}. Experimental ETPA has been studied most extensively in molecular systems, where early fluorescence- and transmission-based measurements reported cross sections as large as $\sigma_e \sim 10^{-18}$--$10^{-17}$~cm$^2$, suggesting a dramatic quantum advantage over classical excitation~\cite{lee2006,varnavski2017,villabona2017}. However, a series of recent experiments with carefully controlled systematics have placed upper bounds on $\sigma_e$ that are orders of magnitude smaller, with many studies detecting no ETPA signal at all~\cite{landes2021,parzuchowski2021,he2024}. Several single-photon mechanisms that can mimic the linear flux dependence expected of ETPA have since been identified, including hot-band absorption from thermally populated vibrational levels~\cite{mikhaylov2022}, single-photon scattering~\cite{hickam2022}, and linear optical losses in transmission geometries~\cite{trianaarango2024}. Because rate scaling alone cannot distinguish ETPA from these artifacts, the controversy motivates a search for qualitative entanglement signatures that have no classical counterpart. Such signatures have been explored theoretically, including time-energy entanglement that imposes pathway-selective interference in multilevel systems~\cite{fei1997,muthukrishnan2004,schlawin2013nc,oka2020} and reveals classically dark bipolariton states in cavity-coupled molecular ensembles~\cite{gu2020}.

Monolayer transition metal dichalcogenides (TMDs) provide an attractive material platform for realizing such a qualitative signature, because their chiral optical selection rules directly couple photon polarization to the valley degree of freedom of excited carriers. The valley structure originates from the inequivalent $K$ and $K'$ points of the hexagonal Brillouin zone, where circularly polarized light selectively couples to interband transitions in a single valley~\cite{xiao2012,mak2012,zeng2012,cao2012}. These materials host tightly bound excitons with binding energies of several hundred meV, arising from reduced dielectric screening and strong Coulomb interaction in two dimensions~\cite{mak2010,chernikov2014,wang2018}. The chiral optical selection rule, combined with spin-valley locking due to the strong spin-orbit interaction, has enabled extensive studies of valley polarization, valley coherence, and valleytronic device concepts~\cite{xu2014,schaibley2016,yu2015}.

At the single-photon level, this polarization-to-valley coupling has been directly demonstrated. Tokman et al.~\cite{tokman2015} showed that absorbing a linearly polarized photon, a coherent superposition of $\sigma^+$ and $\sigma^-$ in the circular basis, produces valley-entangled excitons in a TMD monolayer, with the entanglement verifiable through the polarization of the photoluminescence. This confirms that the polarization structure of quantum light can be coherently imprinted onto the valley degree of freedom.

Beyond single excitons, TMD monolayers support biexcitons (bound states of two excitons) with binding energies of order 20~meV~\cite{you2015,hao2017,barbone2018,chen2018}. The biexciton spectrum is richer than that of the exciton, because electron--hole exchange interaction couples the six four-particle configurations that can be formed from excitons in the $K$ and $K'$ valleys into eigenstates with distinct symmetry and radiative properties~\cite{steinhoff2018}. Three eigenstates are optically bright and three are dark. Among the dark eigenstates, one ($\Phi_3$) is especially relevant because it has nonzero overlap with the optically accessible cross-circular configurations, yet its cascade emission vanishes by destructive interference. This darkness originates not from a spin selection rule, as for the spin-forbidden dark exciton~\cite{zhang2017,molas2017,wang2017}, but from the antisymmetric superposition structure of the eigenstate in configuration space. The fine structure was confirmed by ultrafast pump--probe spectroscopy~\cite{steinhoff2018}. Subsequent theoretical work has further explored the role of exchange in TMD exciton--exciton interactions~\cite{erkensten2021,shahnazaryan2017} and biexciton--polariton formation~\cite{kudlis2025}. The question we address is whether the polarization-to-valley principle extends to two-photon absorption of an entangled photon pair, where the exchange-split biexciton eigenstate structure opens qualitatively new possibilities for state-selective control through the biphoton polarization phase. To our knowledge, no prior work has combined biphoton polarization entanglement with the exchange-split biexciton manifold of a TMD monolayer.

\begin{figure}[t]
	\includegraphics[width=0.8\columnwidth]{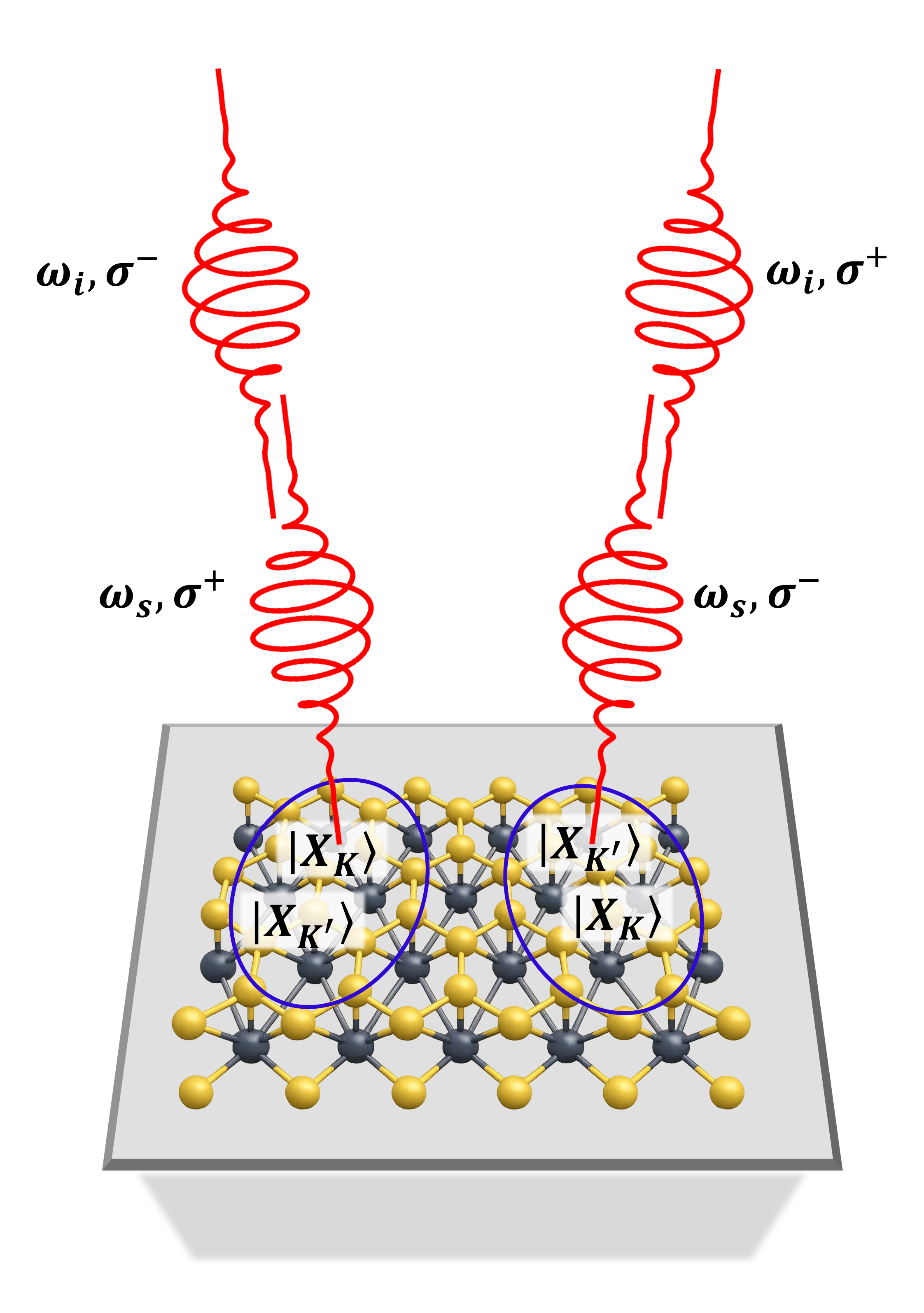}
	\caption{\label{fig:scheme}%
	Cross-circular two-photon excitation scheme in monolayer WSe$_2$. A polarization-entangled photon pair drives two valley pathways: $(\sigma^+,\sigma^-)$ creates $\ket{X_K}$ then $\ket{X_{K'}}$ (left), while $(\sigma^-,\sigma^+)$ creates $\ket{X_{K'}}$ then $\ket{X_K}$ (right).}
\end{figure}

In this work, we propose such a symmetry-based ETPA observable, realized through the interplay between biphoton polarization entanglement and electron--hole exchange symmetry in the biexciton manifold of monolayer WSe$_2$. We consider frequency-nondegenerate cross-circular two-photon excitation through a real intermediate $A$ exciton, as illustrated schematically in Fig.~\ref{fig:scheme}. This sequential absorption through a real intermediate state follows the same excitation principle demonstrated with entangled photon pairs in atomic cesium~\cite{georgiades1995} and rubidium~\cite{dayan2004}. We extend this principle to a solid-state setting. Combining the valley optical selection rules with the exchange-split biexciton eigenstates of Steinhoff et al.~\cite{steinhoff2018} gives eigenstate-resolved pumping rates and Bell-phase selection rules for the bright and dark eigenstate sectors. The dark-eigenstate pumping rate of any separable source obeys an upper bound. No separable polarization state reproduces the entangled eigenstate distribution. We assess the robustness of the selection rule against valley dephasing, intervalley scattering, and imperfections in the photon-pair source at cryogenic temperatures.

\section{\label{sec:theory}Theory: From Entangled Photon Pair to Eigenstate Population}

\subsection{\label{sec:valley}Valley selection rules and two-photon amplitude}

The total Hamiltonian for the coupled light-matter system is
\begin{equation}
	\label{eq:1}
	H = H_\mathrm{mat} + H_\mathrm{field} + V,
\end{equation}
where $H_\mathrm{mat}$ is the material Hamiltonian with eigenstates $\{\ket{g}, \ket{X_K}, \ket{X_{K'}}, \ket{B_j}\}$. Here $\ket{g}$ is the ground state, $\ket{X_K}$ and $\ket{X_{K'}}$ are the bright $A$-excitons in the $K$ and $K'$ valleys (created by $\sigma^+$ and $\sigma^-$ photons, respectively), and $\ket{B_j}$ ($j = 1,\ldots,6$) are the six zero-momentum biexciton configurations classified by valley occupancy and spin pairing of the two constituent excitons~\cite{steinhoff2018} (Table~\ref{tab:eigenstates}); $H_\mathrm{field}$ is the free electromagnetic field Hamiltonian; and $V$ is the electric-dipole interaction in the rotating-wave approximation:
\begin{equation}
	\label{eq:2}
	V = -\hat{\mathbf{d}} \cdot \hat{\mathbf{E}}^{(+)}(t) + \mathrm{h.c.}
\end{equation}
The positive-frequency electric field operator is decomposed into signal ($s$) and idler ($i$) modes with polarization $\sigma$~\cite{shih2021}:
\begin{equation}
	\label{eq:3}
	\begin{split}
		\hat{E}^{(+)}_\sigma(t) &= \sum_{m \in \{s,i\}} \int d\omega\, \mathcal{E}_m(\omega)\hat{a}_{m,\sigma}(\omega)\, e^{-i\omega t},
	\end{split}
\end{equation}
where $\mathcal{E}_m(\omega)$ is the mode-envelope function and $\hat{a}_{m,\sigma}(\omega)$ annihilates a photon in mode $m$ with polarization $\sigma$.

The biexciton is created by the absorption of two photons from the entangled pair. In the interaction picture with $H_0 = H_\mathrm{mat} + H_\mathrm{field}$, the time-evolution operator to second order in $V$ is
\begin{equation}
	\label{eq:4}
	\begin{split}
		U_I(t, -\infty) &= \mathbb{I} + \left(-\frac{i}{\hbar}\right)\int_{-\infty}^{t} dt_1 V_I(t_1) \\
		&\quad + \left(-\frac{i}{\hbar}\right)^2\int_{-\infty}^{t} dt_2 \int_{-\infty}^{t_2} dt_1 \\
		&\qquad\times V_I(t_2) V_I(t_1) + \cdots,
	\end{split}
\end{equation}
where $V_I(t) = e^{iH_0 t/\hbar} V e^{-iH_0 t/\hbar}$ and the time-ordering constraint $t_1 < t_2$ enforces causality. The second order is the leading contribution that connects the ground state to the biexciton manifold. Projecting the second-order term onto the final state $\ket{B_j} \otimes \ket{0}$ (biexciton with photon vacuum), the transition amplitude from the initial state $\ket{g} \otimes \ket{\psi_\mathrm{2ph}}$ is
\begin{equation}
	\label{eq:5}
	\begin{split}
		\mathcal{M}_j &= \left(-\frac{i}{\hbar}\right)^2 \int_{-\infty}^{t} dt_2 \int_{-\infty}^{t_2} dt_1 \\
		&\quad\times \bra{B_j}\bra{0} V_I(t_2) V_I(t_1) \ket{g}\ket{\psi_\mathrm{2ph}},
	\end{split}
\end{equation}
where the two insertions of $V_I$ account for the sequential absorption of two photons.

\begin{figure}[t]
	\includegraphics[width=\columnwidth]{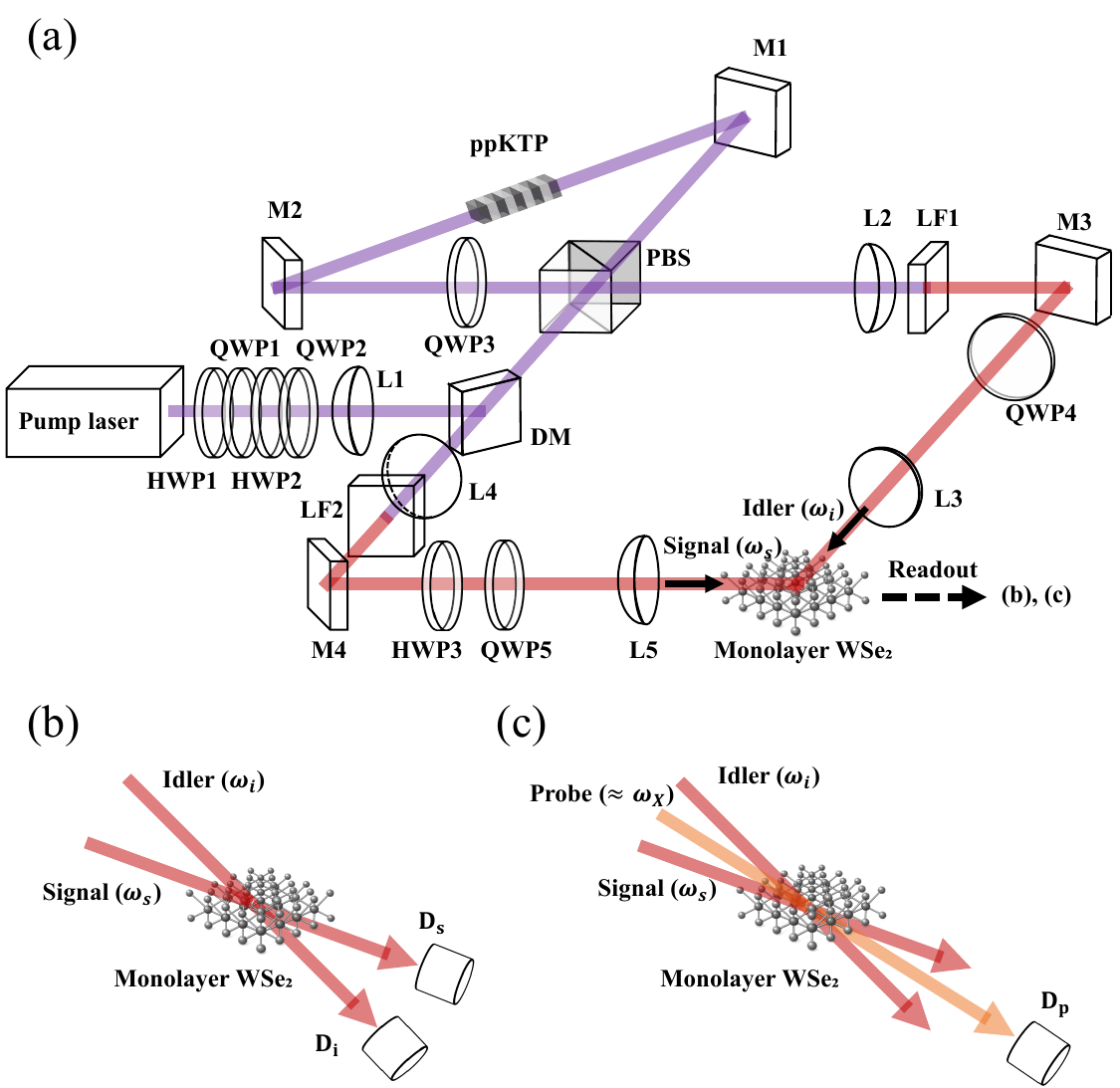}
	\caption{\label{fig:setup}%
	Proposed experimental setup and readout schemes. (a)~Source and sample. A pump laser drives nondegenerate type-II SPDC in a Sagnac interferometer. HWP1 in the pump path sets the Bell-state phase $\varphi$. The DM and LFs remove the residual pump. Wave plates in each arm (QWP4 in one arm and HWP3\,+\,QWP5 in the other) convert the $H$--$V$ output to the cross-circular basis. (b)~Coincidence-transmission readout. The transmitted signal and idler are detected in coincidence at $\mathrm{D}_s$ and $\mathrm{D}_i$. (c)~Pump-probe readout. The SPDC pair acts as the pump, and a weak probe near the $A$-exciton resonance, detected at $\mathrm{D}_p$, monitors the pump-induced change in the probe transmission.
	PBS: polarizing beam splitter; DM: dichroic mirror;
	HWP1--HWP3: half-wave plates; QWP1--QWP5: quarter-wave plates;
	LF1, LF2: long-pass filters; L1--L5: lenses; M1--M4: mirrors;
	$\mathrm{D}_s$ and $\mathrm{D}_i$: single-photon detectors for the signal and idler arms.
	$\mathrm{D}_p$: photodetector for the probe arm.}
\end{figure}

The input field state, $\ket{\psi_\mathrm{2ph}}$, is a broadband polarization-entangled photon pair generated by nondegenerate type-II SPDC in a polarization-based Sagnac interferometer~\cite{kim2006,fedrizzi2007,tischler2016}, shown schematically in Fig.~\ref{fig:setup}(a), with a periodically poled KTP crystal. We assume $\langle n\rangle\ll1$ pair per mode. The Sagnac output is $\ket{\psi_\text{lin}} = (\ket{H_s V_i} + e^{i\varphi}\ket{V_s H_i})/\sqrt{2}$, where the Bell-state phase $\varphi = 4\theta_\mathrm{HWP1}$ is tuned by a half-wave plate in the pump beam~\cite{tischler2016}. The nondegenerate signal and idler wavelengths are set by the phase-matching condition and tuned via the crystal temperature. We assume that the signal photon at frequency $\omega_s = E_X/\hbar$ ($\lambda_s \approx 713$ nm) is resonant with the $A$-exciton transition energy $E_X \approx 1.74$ eV~\cite{wang2018}. The idler frequency is set by the target biexciton eigenstate. For the exchange-split fine structure of monolayer WSe$_2$~\cite{steinhoff2018}, the binding energies span from $E_B^{(\Phi_1)} \approx 19.8$ meV (singlet, most strongly bound) through the triplet pair $E_B^{(\Phi_2)} \approx 12.3$ meV and $E_B^{(\Phi_3)} \approx 11.6$ meV, to $E_B^{(\Phi_6)} \approx 3.1$ meV (see Table~\ref{tab:eigenstates}). Setting the fixed continuous-wave (CW) pump frequency to $\omega_p = \omega_s + \omega_i$ with $\omega_i = (E_X - E_B^{(\Phi_\alpha)})/\hbar$ places the sum frequency on resonance with eigenstate $\Phi_\alpha$. For example, targeting $\Phi_3$ with $E_B \approx 11.6$ meV gives $\lambda_i \approx 718$ nm; targeting $\Phi_1$ with $E_B \approx 19.8$ meV gives $\lambda_i \approx 721$ nm. A dichroic mirror and long-pass filters remove the residual pump beam. Because a passive basis change from $H\text{--}V$ to circular polarization would yield both cross- and cocircular components, wave plates in each arm (HWP3\,+\,QWP5 in one arm and QWP4 in the other) implement the local maps $H \to \sigma^+$, $V \to \sigma^-$, with the HWP3 orientation selecting the cross-circular configuration~\cite{tischler2016}. This yields the cross-circular Bell state:
\begin{equation}
	\label{eq:6}
	\begin{split}
		\ket{\psi_\mathrm{2ph}} &= \frac{1}{\sqrt{2}}\iint d\omega_s d\omega_i f(\omega_s,\omega_i) \\
		&\quad\times\Big[\hat{a}^\dagger_{s,\sigma^+}(\omega_s) \hat{a}^\dagger_{i,\sigma^-}(\omega_i) \\
		&\quad+ e^{i\varphi}\hat{a}^\dagger_{s,\sigma^-}(\omega_s) \hat{a}^\dagger_{i,\sigma^+}(\omega_i)\Big]\ket{0},
	\end{split}
\end{equation}
where $f(\omega_s,\omega_i)$ is the joint spectral amplitude (JSA) encoding the energy-time correlations of the SPDC output. Setting $\varphi = 0$ yields $\ket{\psi^+}$ and $\varphi = \pi$ yields $\ket{\psi^-}$.

The phase dependence of the pair absorption is measured with two readouts. The coincidence readout counts the transmitted signal--idler pairs in coincidence [Fig.~\ref{fig:setup}(b)]. The pump-probe readout records the transmission of a weak probe [Fig.~\ref{fig:setup}(c)]. Section~\ref{sec:experimental} gives the corresponding detection conditions.

The valley optical selection rules in monolayer WSe$_2$ dictate which intermediate exciton is created by each circular polarization~\cite{xiao2012,mak2012,zeng2012,cao2012}:
\begin{equation}
	\label{eq:7}
	\sigma^+ \longrightarrow \ket{X_K}, \quad \sigma^- \longrightarrow \ket{X_{K'}}.
\end{equation}
Explicitly, the nonvanishing dipole matrix elements are
\begin{equation}
	\label{eq:8}
	\bra{X_K}\hat{d} \cdot \mathbf{e}_+\ket{g} = d_0, \quad \bra{X_{K'}}\hat{d} \cdot \mathbf{e}_-\ket{g} = d_0,
\end{equation}
with all cross-valley couplings vanishing ($\bra{X_K}\hat{d} \cdot \mathbf{e}_-\ket{g} = \bra{X_{K'}}\hat{d} \cdot \mathbf{e}_+\ket{g} = 0$). Here $\mathbf{e}_\pm = \mp(\mathbf{e}_x \pm i\mathbf{e}_y)/\sqrt{2}$ are the circular polarization unit vectors and $d_0$ is the ground-to-exciton transition dipole matrix element.

For the cross-circular components of the Bell state in Eq.~\eqref{eq:6}, these selection rules constrain the two-photon pathways to the opposite-valley configurations~\cite{steinhoff2018}:
\begin{subequations}
	\label{eq:9}
	\begin{align}
		&(\sigma^+, \sigma^-): \quad \ket{g} \xrightarrow{\omega_s, \sigma^+} \ket{X_K} \xrightarrow{\omega_i, \sigma^-} \ket{B_2}, \label{eq:9a} \\
		&(\sigma^-, \sigma^+): \quad \ket{g} \xrightarrow{\omega_s, \sigma^-} \ket{X_{K'}} \xrightarrow{\omega_i, \sigma^+} \ket{B_3}, \label{eq:9b}
	\end{align}
\end{subequations}
where $\ket{B_2}$ and $\ket{B_3}$ are biexciton configurations composed of excitons in the $(K, K')$ and $(K', K)$ valley pairs, respectively. These are two of the six zero-momentum biexciton configurations classified by Steinhoff et al.~\cite{steinhoff2018}; the remaining four are inaccessible to cross-circular excitation (Table~\ref{tab:eigenstates} lists all six configurations and their eigenstate decomposition). Each polarization sequence activates exactly one valley channel through exactly one intermediate state. This one-to-one mapping determines the eigenstate selectivity.

To evaluate the Dyson series amplitude Eq.~\eqref{eq:5}, we insert a complete set of intermediate one-exciton states $\sum_n \ket{n}\bra{n}$ between the two interactions. The matter matrix element becomes $\bra{B_j}\hat{d}_{\sigma_2}\ket{n}\bra{n}\hat{d}_{\sigma_1}\ket{g}$ (where subscripts 1 and 2 label the temporal vertex ordering, $t_1 < t_2$), which selects a specific valley intermediate $\ket{n}$ via the selection rules Eq.~\eqref{eq:7}. The field part requires contracting the two annihilation operators $\hat{a}_{m_2,\sigma_2}(\omega_2)\hat{a}_{m_1,\sigma_1}(\omega_1)$ with the two creation operators $\hat{a}^\dagger_{s,\mu}(\omega_s)\hat{a}^\dagger_{i,\nu}(\omega_i)$ in $\ket{\psi_\mathrm{2ph}}$ [Eq.~\eqref{eq:6}]. Since signal and idler occupy distinct modes with $[\hat{a}_{s,\mu}(\omega), \hat{a}^\dagger_{i,\nu}(\omega')] = 0$, there are exactly two ways to pair the annihilation operators with the creation operators.

The later interaction ($t_2$) annihilates the signal photon and the earlier interaction ($t_1$) annihilates the idler. The field matrix element evaluates to
\begin{equation}
	\label{eq:10}
	\bra{0}\hat{a}_{s,\sigma_2}(\omega_s)\hat{a}_{i,\sigma_1}(\omega_i) \ket{\psi_\mathrm{2ph}} = \Psi(\omega_s, \sigma_2; \omega_i, \sigma_1),
\end{equation}
which pairs the later vertex polarization $\sigma_2$ with the signal label and $\sigma_1$ with the idler label -- Contraction A ($m_2 = s$, $m_1 = i$). The intermediate-state propagator is evaluated at the idler frequency $\omega_i$.

The later interaction annihilates the idler photon and the earlier interaction annihilates the signal. The field matrix element evaluates to
\begin{equation}
	\label{eq:11}
	\bra{0}\hat{a}_{i,\sigma_2}(\omega_i)\hat{a}_{s,\sigma_1}(\omega_s) \ket{\psi_\mathrm{2ph}} = \Psi(\omega_s, \sigma_1; \omega_i, \sigma_2),
\end{equation}
with the polarization indices swapped relative to Contraction A -- thus, Contraction B ($m_2 = i$, $m_1 = s$). The intermediate-state propagator is now evaluated at the signal frequency $\omega_s$.

Evaluating the time-ordered integral in Eq.~\eqref{eq:5} (Appendix~\ref{sec:appA}), each contraction yields a retarded intermediate-state propagator and a common final-state Lorentzian $L(\omega_s + \omega_i) = [(\omega_{B_j} - \omega_s - \omega_i) - i\gamma_B]^{-1}$, where $\omega_{B_j} \equiv (E_{B_j} - E_g)/\hbar$ is the biexciton transition frequency from the ground state, enforcing two-photon energy conservation.

Defining the frequency-dependent two-photon matrix element
\begin{equation}
	\label{eq:12}
	T_j(\omega) = \sum_n \frac{\bra{B_j}\hat{d}_{\sigma_2}\ket{n} \bra{n}\hat{d}_{\sigma_1}\ket{g}} {\hbar(\omega_{ng} - \omega) - i\hbar\gamma_n},
\end{equation}
the total amplitude for populating configuration $\ket{B_j}$ takes the form (see Appendix~\ref{sec:appA})
\begin{equation}
	\label{eq:13}
	\begin{split}
		\mathcal{M}_j &= \iint d\omega_s d\omega_i \Big[\Psi(\omega_s, \sigma_2; \omega_i, \sigma_1) T_j(\omega_i) \\
		&\quad + \Psi(\omega_s, \sigma_1; \omega_i, \sigma_2) T_j(\omega_s)\Big] L(\omega_s{+}\omega_i),
	\end{split}
\end{equation}
where $\Psi(\omega_s, \sigma; \omega_i, \sigma')$ is the biphoton effective wavefunction for polarization pair $(\sigma, \sigma')$. Each term is weighted by a different joint spectral amplitude component, reflecting the polarization label of whichever photon is absorbed at each vertex.

Since only cross-circular polarization pairs contribute to the two-photon pathways [Eqs.~\eqref{eq:9a} and~\eqref{eq:9b}], the relevant components are those with $(\mu,\nu) \in \{(\sigma^+,\sigma^-), (\sigma^-,\sigma^+)\}$. Evaluating these from the explicit biphoton state [Eqs.~\eqref{eq:6}, \eqref{eq:10}, and~\eqref{eq:11}]:
\begin{subequations}
	\label{eq:14}
	\begin{align}
		\Psi(\omega_s, \sigma^+; \omega_i, \sigma^-) &= \frac{f(\omega_s,\omega_i)}{\sqrt{2}}, \label{eq:14a}\\
		\Psi(\omega_s, \sigma^-; \omega_i, \sigma^+) &= \frac{e^{i\varphi} f(\omega_s,\omega_i)}{\sqrt{2}}. \label{eq:14b}
	\end{align}
\end{subequations}
The two components satisfy
\begin{equation}
	\label{eq:15}
	\Psi(\omega_s, \sigma^-; \omega_i, \sigma^+) = e^{i\varphi}\, \Psi(\omega_s, \sigma^+; \omega_i, \sigma^-).
\end{equation}
For $\varphi = 0$ ($\ket{\psi^+}$), the two components are equal; for $\varphi = \pi$ ($\ket{\psi^-}$), they differ by a sign. The phase $e^{i\varphi}$ encodes the polarization correlation of the entangled pair, which is a distinct degree of freedom from the frequency correlations contained in $f$.

Substituting Eq.~\eqref{eq:14} into Eq.~\eqref{eq:13} for the two cross-circular configurations yields
\begin{subequations}
	\label{eq:16}
	\begin{align}
		\mathcal{M}_2 &= \frac{1}{\sqrt{2}}\iint d\omega_s d\omega_i f(\omega_s,\omega_i) \notag\\
		&\quad\times\Big[ e^{i\varphi} T_2(\omega_i) + T_2(\omega_s) \Big] L(\omega_s {+} \omega_i), \label{eq:16a} \\
		\mathcal{M}_3 &= \frac{1}{\sqrt{2}}\iint d\omega_s d\omega_i f(\omega_s,\omega_i) \notag\\
		&\quad\times\Big[ T_3(\omega_i) + e^{i\varphi} T_3(\omega_s) \Big] L(\omega_s {+} \omega_i). \label{eq:16b}
	\end{align}
\end{subequations}
The Bell-state phase $e^{i\varphi}$ appears on the $T_2(\omega_i)$ term in $\mathcal{M}_2$ but on the $T_3(\omega_s)$ term in $\mathcal{M}_3$, reflecting which polarization branch of the Bell state feeds each valley pathway.

\subsection{\label{sec:finestructure}Biexciton fine structure and eigenstate projection}

The electron-hole exchange interaction mixes the six zero-momentum biexciton configurations $\ket{B_1}$-$\ket{B_6}$~\cite{steinhoff2018}. Diagonalizing the configuration-basis Hamiltonian -- which includes nonlocal exchange (coupling constant $U$) and local-field corrections -- yields six eigenstates with distinct decompositions, energies, binding energies, and emission properties.

\begin{table*}[htbp]
	\caption{\label{tab:eigenstates}%
	Biexciton eigenstates from the exchange Hamiltonian~\cite{steinhoff2018}. Single ($\uparrow$/$\downarrow$) and double ($\Uparrow$/$\Downarrow$) arrows denote electron and hole spin projections, respectively, with $\uparrow$/$\Uparrow$ residing at $K$ and $\downarrow$/$\Downarrow$ at $K'$; each four-arrow symbol is ordered as $e_1 h_1 e_2 h_2$ for the two electron--hole pairs composing the configuration. Columns list the expansion coefficients $\braket{\Phi_\alpha | B_j}$ in the configuration basis, the eigenenergy in the single-constant approximation (from Eq.~(8) of Ref.~\onlinecite{steinhoff2018}), the binding energy $E_B$ from the full microscopic calculation for WSe$_2$ on sapphire~\cite{steinhoff2018}, the $B_2 \leftrightarrow B_3$ symmetry class, the cross-circular oscillator strength $|d_\alpha|^2 \propto |\braket{\Phi_\alpha | B_2} + \braket{\Phi_\alpha | B_3}|^2$, and bright/dark character. The $E_B$ values include local-field corrections.}
	\begin{ruledtabular}
		\begin{tabular}{cccccccccccc}
			$\Phi_\alpha$
			& $B_1$ & $B_2$ & $B_3$ & $B_4$ & $B_5$ & $B_6$
			& Energy & $E_B$ (meV) & $B_2\!\leftrightarrow\!B_3$ & $|d_\alpha|^2$ & Character \\
			& $\uparrow\Uparrow\uparrow\Uparrow$
			& $\uparrow\Uparrow\downarrow\Downarrow$
			& $\downarrow\Downarrow\uparrow\Uparrow$
			& $\uparrow\Downarrow\downarrow\Uparrow$
			& $\downarrow\Uparrow\uparrow\Downarrow$
			& $\downarrow\Downarrow\downarrow\Downarrow$
			& & & & & \\
			\colrule
			$\Phi_1$ & $1/\sqrt{6}$ & $-1/\sqrt{6}$ & $-1/\sqrt{6}$ & $-1/\sqrt{6}$ & $-1/\sqrt{6}$ & $1/\sqrt{6}$ & $E$ & 19.8 & Sym. & 2/3 & Bright \\
			$\Phi_2$ & $0$ & $1/2$ & $1/2$ & $-1/2$ & $-1/2$ & $0$ & $E + 2U$ & 12.3 & Sym. & 1 & Bright \\
			$\Phi_{3}$ & $0$ & $1/\sqrt{2}$ & $-1/\sqrt{2}$ & $0$ & $0$ & $0$ & $E + 2U$ & 11.6 & Anti. & 0 & Dark \\
			$\Phi_4$ & $0$ & $0$ & $0$ & $1/\sqrt{2}$ & $-1/\sqrt{2}$ & $0$ & $E + 2U$ & 11.6 & --- & --- & Dark \\
			$\Phi_5$ & $1/\sqrt{2}$ & $0$ & $0$ & $0$ & $0$ & $-1/\sqrt{2}$ & $E + 4U$ & 3.4 & --- & --- & Dark \\
			$\Phi_6$ & $2/\sqrt{12}$ & $1/\sqrt{12}$ & $1/\sqrt{12}$ & $1/\sqrt{12}$ & $1/\sqrt{12}$ & $2/\sqrt{12}$ & $E + 6U$ & 3.1 & Sym. & 1/3 & Bright \\
		\end{tabular}
	\end{ruledtabular}
\end{table*}

Of the six eigenstates, four ($\Phi_1$, $\Phi_2$, $\Phi_3$, $\Phi_6$) have nonzero overlap with the optically accessible cross-circular configurations $\ket{B_2}$ and $\ket{B_3}$; the remaining two ($\Phi_4$, $\Phi_5$) are inaccessible in this sector. Table~\ref{tab:eigenstates} shows that all three bright eigenstates have equal overlaps with $\ket{B_2}$ and $\ket{B_3}$ and are symmetric under $B_2 \leftrightarrow B_3$. The state $\Phi_{3}$ is the unique antisymmetric eigenstate in the full six-dimensional Hilbert space. Its vanishing oscillator strength follows directly, $d_3 \propto \braket{\Phi_{3} | B_2 } + \braket{\Phi_{3} | B_3 } = 0$.

\subsubsection{Eigenstate decomposition of the cross-circular sector}\label{sec:eigendecomp}

The symmetric and antisymmetric superpositions of $\ket{B_2}$ and $\ket{B_3}$ have distinct eigenstate decompositions (Table~\ref{tab:eigenstates}):
\begin{subequations}
	\label{eq:17}
	\begin{equation}
		\label{eq:17a}
		\frac{1}{\sqrt{2}}(\ket{B_2} + \ket{B_3})=-\frac{1}{\sqrt{3}}\ket{\Phi_1} + \frac{1}{\sqrt{2}}\ket{\Phi_2} + \frac{1}{\sqrt{6}}\ket{\Phi_6},
	\end{equation}
	\begin{equation}
		\label{eq:17b}
		\frac{1}{\sqrt{2}}(\ket{B_2} - \ket{B_3}) = \ket{\Phi_{3}}.
	\end{equation}
\end{subequations}
The symmetric superposition has nonzero overlap with all three bright eigenstates; the antisymmetric superposition is exactly the dark eigenstate $\Phi_3$. These overlaps determine which eigenstates are accessible for a given Bell phase and set the geometric coupling efficiency for each eigenstate, but the absolute excitation rate also depends on the spectral factor $\kappa_\alpha$ (Sec.~\ref{sec:pumping}).

\subsubsection{Separation of the Bell-state phase}\label{sec:exactfactor}

Time-reversal symmetry of the TMD band structure at zero magnetic field guarantees identical dipole matrix elements and dephasing rates in the two valleys ($d_0^{(K)} = d_0^{(K')}$, $\gamma_K = \gamma_{K'}$), so that the two-photon kernels satisfy $T_2(\omega) = T_3(\omega) \equiv T(\omega)$. Since the eigenstate decomposition [Eqs.~\eqref{eq:17a} and~\eqref{eq:17b}] shows that $\mathcal{M}_2 + \mathcal{M}_3$ controls the bright eigenstates ($\Phi_1$, $\Phi_2$, $\Phi_6$) while $\mathcal{M}_2 - \mathcal{M}_3$ controls the dark eigenstate $\Phi_3$, we form these combinations from the cross-circular amplitudes [Eqs.~\eqref{eq:16a} and~\eqref{eq:16b}]:
\begin{subequations}
	\label{eq:18}
	\begin{align}
		\mathcal{M}_2 {+} \mathcal{M}_3 &= \frac{1{+}e^{i\varphi}}{\sqrt{2}}\iint d\omega_s d\omega_i f\big[T(\omega_s) {+} T(\omega_i)\big]L, \label{eq:18a}\\
		\mathcal{M}_2 {-} \mathcal{M}_3 &= \frac{1{-}e^{i\varphi}}{\sqrt{2}}\iint d\omega_s d\omega_i f\big[T(\omega_s) {-} T(\omega_i)\big]L, \label{eq:18b}
	\end{align}
\end{subequations}
where $f \equiv f(\omega_s, \omega_i)$ and $L \equiv L(\omega_s + \omega_i)$. The prefactors $(1 \pm e^{i\varphi})$ are independent of the spectral integrals; the Bell-state phase factorizes exactly from the material response.

For the symmetric Bell state $\ket{\psi^+}$ ($\varphi = 0$), the antisymmetric combination vanishes, $\mathcal{M}_2 - \mathcal{M}_3 = 0$, so the dark eigenstate $\Phi_3$, which requires precisely this combination Eq.~\eqref{eq:17b}, receives zero population. The excitation is then confined to the bright sector.

For the antisymmetric Bell state $\ket{\psi^-}$ ($\varphi = \pi$), the symmetric combination vanishes, $\mathcal{M}_2 + \mathcal{M}_3 = 0$, so all bright eigenstates, which require the symmetric combination Eq.~\eqref{eq:17a}, receive zero population. The prepared state is then the antisymmetric eigenstate $\ket{\Phi_{3}}$.

The phase-dependent extinctions rely on two independent conditions. Valley symmetry, $T_2(\omega)=T_3(\omega)$, makes the two cross-circular pathways share the same material response function. The common-JSA assumption of Eq.~\eqref{eq:6} gives the two Bell components the same spectral amplitude. A magnetic field, magnetic disorder, or photoinduced valley polarization can make the $K$ and $K'$ responses unequal, $T_2\neq T_3$. In contrast, nonmagnetic strain or dielectric disorder that shifts both valley responses together changes the common lineshape without lifting the bright-state extinction at $\varphi=\pi$ or the dark-state extinction at $\varphi=0$. Polarization-dependent spectral distortion in the beam path, such as chromatic birefringence, gives the two Bell components different JSAs.

\subsubsection{Dominant-contraction approximation}\label{sec:dominant}

A further simplification arises from the nondegenerate excitation scheme. Because the signal and idler central frequencies are separated by the binding energy $E_B^{(\Phi_\alpha)}$ $\gg \hbar\gamma_X \approx 1.5$ meV~\cite{moody2015}, only one time-ordering has a resonant intermediate state. The suppression ratio is
\begin{equation}
	\label{eq:19}
	\frac{|T(\omega_i)|^2}{|T(\omega_s)|^2} = \frac{\hbar^2 \gamma_X^2}{(E_B^{(\Phi_\alpha)})^2 + \hbar^2\gamma_X^2} \approx \left(\frac{\hbar \gamma_X}{E_B^{(\Phi_\alpha)}}\right)^{2},
\end{equation}
which is $\approx0.015$ for $\ket{\Phi_2}$ and $\ket{\Phi_3}$ ($E_B\approx12$~meV) and $\approx0.006$ for $\ket{\Phi_1}$ ($E_B\approx19.8$~meV), so the off-resonant time ordering is suppressed by roughly two orders of magnitude in rate for these states, while for the weakly bound $\ket{\Phi_6}$ ($E_B=3.1$~meV) the ratio reaches ${\sim}0.19$. The dominant-contraction approximation is therefore less accurate for $\ket{\Phi_6}$. Dropping the off-resonant contraction ($T_2(\omega_i)$ in Eq.~\eqref{eq:16a} and $T_3(\omega_i)$ in Eq.~\eqref{eq:16b}) gives
\begin{subequations}
	\label{eq:20}
	\begin{align}
		\mathcal{M}_2 &\approx \frac{1}{\sqrt{2}}\iint d\omega_s\, d\omega_i\; f\, T_2(\omega_s)\, L, \label{eq:20a}\\
		\mathcal{M}_3 &\approx \frac{e^{i\varphi}}{\sqrt{2}}\iint d\omega_s\, d\omega_i\; f\, T_3(\omega_s)\, L. \label{eq:20b}
	\end{align}
\end{subequations}
The Bell-state phase $e^{i\varphi}$ remains only in $\mathcal{M}_3$, since in $\mathcal{M}_2$ it entered through the off-resonant contraction $T_2(\omega_i)$, which has been suppressed. Valley symmetry ($T_2 = T_3$, Sec.~\ref{sec:exactfactor}) then yields the simple ratio
\begin{equation}
	\label{eq:21}
	\mathcal{M}_3 \approx e^{i\varphi}\, \mathcal{M}_2,
\end{equation}
valid to corrections of order $\hbar\gamma_X/E_B^{(\Phi_\alpha)}$ in amplitude.

Since cross-circular excitation can only populate $\ket{B_2}$ and $\ket{B_3}$ [Eqs.~\eqref{eq:9a} and~\eqref{eq:9b}], the second-order perturbation amplitude Eq.~\eqref{eq:5} produces a biexciton state proportional to the sum of configuration amplitudes:
\begin{equation}
	\label{eq:22}
	\ket{\psi_\mathrm{BX}} \propto \mathcal{M}_2\ket{B_2} + \mathcal{M}_3\ket{B_3}.
\end{equation}
Substituting Eq.~\eqref{eq:21}, the common scalar amplitude $\mathcal{M}_2$ factors out:
\begin{equation}
	\label{eq:23}
	\ket{\psi_\mathrm{BX}} \propto \mathcal{M}_2\big(\ket{B_2} + e^{i\varphi}\ket{B_3}\big).
\end{equation}
Because $\mathcal{M}_2$ is an overall complex number, it does not affect the quantum state and drops out upon normalization. The biexciton configurations are orthogonal, $\braket{B_j | B_k} = \delta_{jk}$, so $\|\ket{B_2} + e^{i\varphi}\ket{B_3}\|^2 = 1 + |e^{i\varphi}|^2 = 2$. The normalized prepared state is therefore
\begin{equation}
	\label{eq:24}
	\ket{\psi_\mathrm{prep}} = \frac{1}{\sqrt{2}}\big(\ket{B_2} + e^{i\varphi}\ket{B_3}\big),
\end{equation}
which interpolates continuously between the symmetric ($\varphi = 0$) and antisymmetric ($\varphi = \pi$) limits of Eqs.~\eqref{eq:17a} and~\eqref{eq:17b}. This state encodes the relative phase and amplitude between valley pathways as set by the Bell-state phase.

Within the dominant-contraction limit, valley symmetry guarantees that the $K$ and $K'$ pathways have identical material response ($T_2(\omega) = T_3(\omega)$). As a result, for the sinc joint spectral amplitude of type-II SPDC at the resonant operating point, both configuration amplitudes share a common resonant two-photon amplitude $\mathcal{T}$ (see Appendix~\ref{sec:appB}):
\begin{equation}
	\label{eq:25}
	\mathcal{M}_2 = \frac{2\pi\,\mathcal{N}\, L(\omega_p)}{\sqrt{2}\,T_e}\,\mathcal{T}, \qquad \mathcal{M}_3 = \frac{2\pi\,e^{i\varphi}\,\mathcal{N}\, L(\omega_p)}{\sqrt{2}\,T_e}\,\mathcal{T},
\end{equation}
where $\mathcal{N}$ is a source spectral-amplitude prefactor, $L(\omega_p)$ is the final-state Lorentzian, and
\begin{equation}
	\label{eq:26}
	\mathcal{T} = \frac{d_0 \cdot d_{XX,X}}{-i\hbar\gamma_X}\left(1 - e^{-\gamma_X T_e/2}\right).
\end{equation}
Here $d_{XX,X}$ is the exciton-to-biexciton transition dipole and $\gamma_X$ is the exciton dephasing rate. $T_e$ is the entanglement time, the width of the window within which the signal and idler photons arrive together at the sample. In a down-conversion source it is set by the crystal length and the group-velocity mismatch, so a shorter $T_e$ corresponds to a broader bandwidth. Equation~\eqref{eq:26} is the result of evaluating the spectral integral in the dominant-contraction amplitude for the physical sinc-shaped JSA of type-II SPDC.

The complete $T_e$ dependence of the transition amplitude Eq.~\eqref{eq:25} resides in the factor $(1 - e^{-\gamma_X T_e/2})/T_e$, which encodes the spectral overlap between the biphoton bandwidth and the exciton resonance. In the broad-bandwidth limit ($\gamma_X T_e \ll 1$), only a small spectral fraction overlaps with the resonance; the overlap factor reduces to $\gamma_X T_e/2$, canceling the $1/(\gamma_X T_e)$ dependence so that the amplitude saturates. In the narrow-bandwidth limit ($\gamma_X T_e \gg 1$), the overlap saturates to unity but the amplitude still decreases as $1/T_e$, so further narrowing only reduces the pumping rate without improving resonance coverage.

\subsection{\label{sec:pumping}Pumping rates}

The transition probability into eigenstate $\ket{\Phi_\alpha}$ is obtained by projecting the prepared biexciton state onto $\ket{\Phi_\alpha}$. For the cross-circular pairs $(\mu,\nu),(\mu',\nu')\in\{(\sigma^+,\sigma^-),(\sigma^-,\sigma^+)\}$, the pumping rate is (see Appendix~\ref{sec:appC} for derivation)
\begin{equation}
	\label{eq:27}
	W_\alpha = \kappa_\alpha \sum_{\mu\nu,\mu'\nu'} \braket{\Phi_\alpha | B_{\mu\nu}} \braket{B_{\mu'\nu'}| \Phi_\alpha} \rho_{\mu\nu,\mu'\nu'},
\end{equation}
where $B_{\sigma^+\sigma^-}=B_2$ and $B_{\sigma^-\sigma^+}=B_3$ [Eqs.~\eqref{eq:9a} and~\eqref{eq:9b}]. The matrix $\rho_{\mu\nu,\mu'\nu'}$ is the polarization density matrix of the input biphoton state. For cocircular input, the pair set is $\{(\sigma^+,\sigma^+),(\sigma^-,\sigma^-)\}$ with $B_{\sigma^+\sigma^+}=B_1$ and $B_{\sigma^-\sigma^-}=B_6$. The eigenstate-dependent prefactor is
\begin{equation}
	\label{eq:28}
	\kappa_\alpha \equiv \kappa_0\, \frac{\gamma_{B}^2} {(\omega_{\Phi_\alpha g}-\omega_p)^2+\gamma_{B}^2},
\end{equation}
where
\begin{equation}
	\label{eq:29}
	\kappa_0 \equiv R_\mathrm{pairs} \frac{2\pi^2\,|\mathcal{N}|^2 |\mathcal{T}|^2}{T_e^2\,\gamma_{B}^2}
\end{equation}
is the on-resonance pumping-rate prefactor (i.e., the value of $\kappa_\alpha$ when $\omega_p = \omega_{\Phi_\alpha g}$), and $R_\mathrm{pairs}$ is the SPDC pair rate. In the low-flux regime, the ETPA rate is linear in $R_\mathrm{pairs}$~\cite{georgiades1995,raymer2021,schlawin2013nc}. The quantity $\mathcal{N}$ is the source spectral-amplitude prefactor [Eq.~\eqref{eq:B1}]. The parameter $\gamma_{B}$ is the biexciton linewidth (half-width at half-maximum).

We now evaluate Eq.~\eqref{eq:27} for the continuously tunable Bell state Eq.~\eqref{eq:6}. Because the spectral degrees of freedom have already been absorbed into the state-dependent prefactors $\kappa_\alpha$, the remaining light-state dependence enters only through the polarization-sector density matrix $\rho_{\mu\nu,\mu'\nu'}$. For a general biphoton polarization state $\ket{\psi}_\mathrm{pol} = \sum_{\mu\nu} c_{\mu\nu}\ket{\mu}_s\ket{\nu}_i$, this matrix has diagonal elements (populations) $p_{\mu\nu} \equiv \bra{\psi} \hat{a}^\dagger_{s,\mu}\hat{a}^\dagger_{i,\nu} \hat{a}_{i,\nu}\hat{a}_{s,\mu}\ket{\psi} = |c_{\mu\nu}|^2$ and off-diagonal elements (coherences) $\rho_{\mu\nu,\mu'\nu'} \equiv \bra{\psi} \hat{a}^\dagger_{s,\mu'}\hat{a}^\dagger_{i,\nu'} \hat{a}_{i,\nu}\hat{a}_{s,\mu}\ket{\psi} = c_{\mu\nu} c_{\mu'\nu'}^*$. For the biphoton state $\ket{\psi_\mathrm{2ph}}$ of Eq.~\eqref{eq:6}, the polarization state is the cross-circular Bell state
\begin{equation}
	\label{eq:30}
	\ket{\psi(\varphi)}_\mathrm{Bell} = \frac{1}{\sqrt{2}}\left(\ket{\sigma^+}_s\ket{\sigma^-}_i + e^{i\varphi}\ket{\sigma^-}_s\ket{\sigma^+}_i\right),
\end{equation}
with polarization density matrix
\begin{equation}
	\label{eq:31}
	\rho_\mathrm{Bell}(\varphi) = \ket{\psi(\varphi)}_\mathrm{Bell}\bra{\psi(\varphi)}_\mathrm{Bell}
\end{equation}
and nonzero matrix elements
\begin{gather*}
	p_{\sigma^+\sigma^-} = p_{\sigma^-\sigma^+} = \tfrac{1}{2}, \quad \rho_{\sigma^+\sigma^-,\sigma^-\sigma^+} = \tfrac{e^{-i\varphi}}{2}.
\end{gather*}
In the cross-circular sector, only the $(\mu,\nu) \in \{(\sigma^+,\sigma^-),(\sigma^-,\sigma^+)\}$ pairs contribute. For each eigenstate, the sum in Eq.~\eqref{eq:27} reduces to four terms:
\begin{equation}
	\label{eq:32}
	\begin{split}
		W_\alpha &= \kappa_\alpha\Big[ |\braket{\Phi_\alpha | B_2}|^2 p_{\sigma^+\sigma^-} + |\braket{\Phi_\alpha | B_3}|^2 p_{\sigma^-\sigma^+} \\
		&\quad + \braket{\Phi_\alpha | B_2}\braket{B_3 | \Phi_\alpha} \rho_{\sigma^+\sigma^-,\sigma^-\sigma^+} \\
		&\quad + \braket{\Phi_\alpha | B_3}\braket{B_2 | \Phi_\alpha} \rho_{\sigma^-\sigma^+,\sigma^+\sigma^-}\Big].
	\end{split}
\end{equation}
For the pure Bell state, this four-term sum simplifies to a compact form. Using $p_{\sigma^+\sigma^-} = p_{\sigma^-\sigma^+} = 1/2$ and $\rho_{\sigma^+\sigma^-,\sigma^-\sigma^+} = e^{-i\varphi}/2$, the four terms in Eq.~\eqref{eq:32} combine into
\begin{equation}
	\label{eq:33}
	W_\alpha(\varphi) = \kappa_\alpha|\braket{\Phi_\alpha |\psi_\mathrm{prep}}|^2,
\end{equation}
where $\ket{\psi_\mathrm{prep}} = (\ket{B_2} + e^{i\varphi}\ket{B_3})/\sqrt{2}$ [Eq.~\eqref{eq:24}]. This step uses the fact that the overlaps $\braket{\Phi_\alpha | B_2}$ and $\braket{\Phi_\alpha | B_3}$ in Table~\ref{tab:eigenstates} are all real (the exchange Hamiltonian in the configuration basis is a real symmetric matrix), so the cross terms in Eq.~\eqref{eq:32} combine straightforwardly with the diagonal terms into the squared modulus. The separation of $W_\alpha$ into a spectral prefactor $\kappa_\alpha$ and a geometric overlap makes the symmetry-based selection rule transparent. The $\varphi$~dependence resides entirely in the overlap, while the spectral factor $\kappa_\alpha$ is phase-independent.

The relevant overlaps are listed in Table~\ref{tab:eigenstates}. For each bright eigenstate $\alpha \in \{1,2,6\}$, the symmetric overlaps $\braket{ \Phi_\alpha | B_2 } = \braket{ \Phi_\alpha | B_3 } \equiv g_\alpha$ (with $g_1 = -1/\sqrt{6}$, $g_2 = 1/2$, $g_6 = 1/\sqrt{12}$) give
\begin{equation}
	\label{eq:34}
	\begin{split}
		W_\alpha(\varphi) &= \kappa_\alpha |g_\alpha|^2\left[p_{\sigma^+\sigma^-} + p_{\sigma^-\sigma^+} + 2\mathrm{Re}(\rho_{\sigma^+\sigma^-,\sigma^-\sigma^+})\right] \\
		&= \kappa_\alpha |g_\alpha|^2(1+\cos\varphi), \quad \alpha \in \{1,2,6\}.
	\end{split}
\end{equation}
Every bright eigenstate individually carries the same $(1 + \cos\varphi)$ modulation; only the overall scale $\kappa_\alpha |g_\alpha|^2$ differs. For the dark eigenstate (the antisymmetric superposition $\ket{B_2} - \ket{B_3}$), the cross term $\braket{ \Phi_{3} | B_2 }\braket{ B_3 | \Phi_{3} } = -1/2$ reverses the coherence contribution, replacing $(1+\cos\varphi)$ with $(1-\cos\varphi)$:
\begin{equation}
	\label{eq:35}
	\begin{split}
		W_3(\varphi) &= \frac{\kappa_3}{2}\left[p_{\sigma^+\sigma^-} + p_{\sigma^-\sigma^+} - 2\mathrm{Re}(\rho_{\sigma^+\sigma^-,\sigma^-\sigma^+})\right] \\
		&= \frac{\kappa_3}{2}(1-\cos\varphi).
	\end{split}
\end{equation}

\begin{figure}[t]
	\includegraphics[width=\columnwidth]{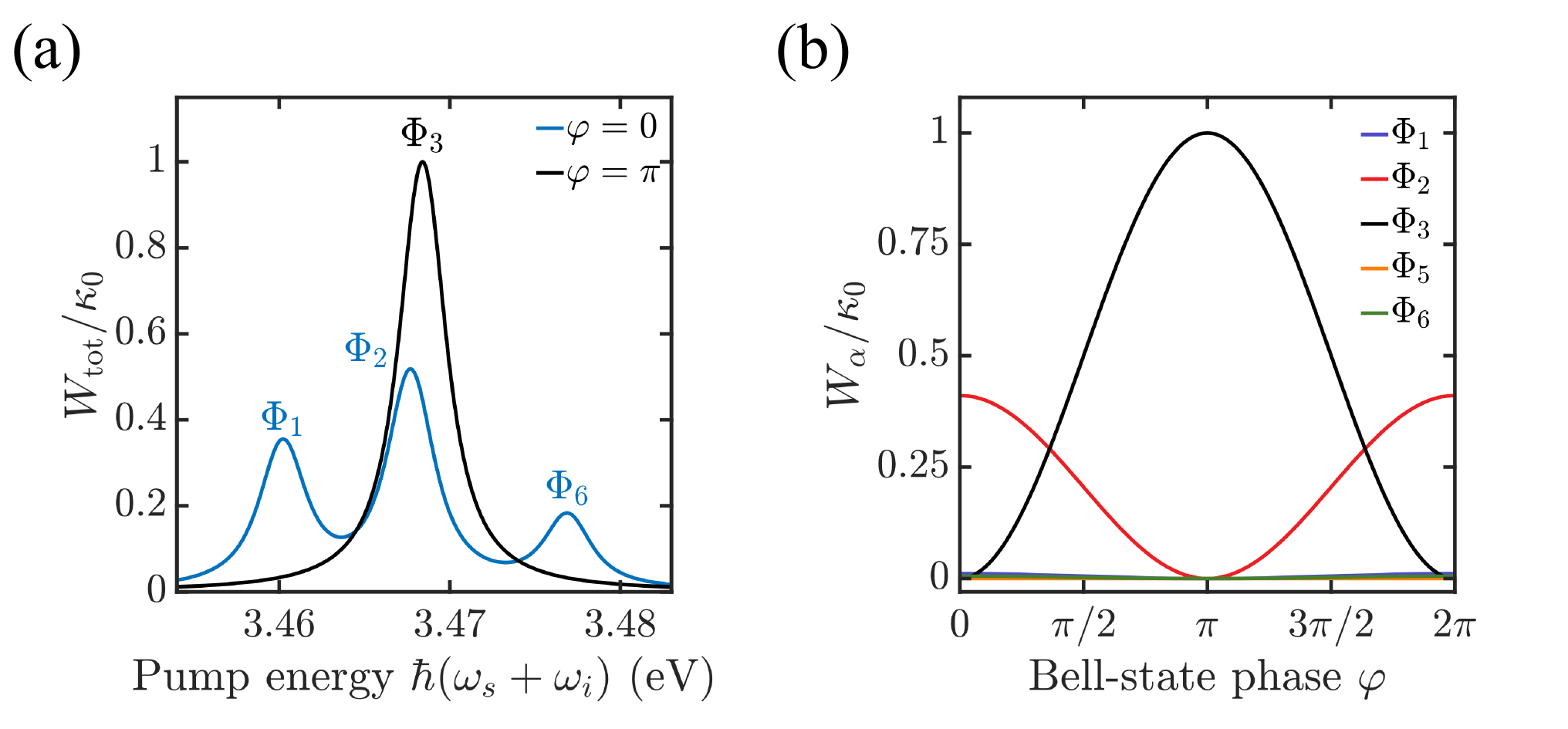}
	\caption{\label{fig:bell}%
	Bell-state pair-absorption rates. (a)~The total pair-absorption rate $W_\mathrm{tot}=\sum_\alpha W_\alpha$ as a function of pump energy, normalized by the resonant spectral factor $\kappa_0$. The labels mark the eigenstate contribution associated with each peak. The homogeneous HWHM is $\hbar\gamma_B=1.5$~meV. Each Lorentzian in Eq.~\eqref{eq:28} is convolved with a Gaussian broadening function of FWHM $\Sigma=1$~meV~\cite{ajayi2017}. At $\varphi=0$, the total spectrum contains the $\Phi_1$, $\Phi_2$, and $\Phi_6$ contributions. At $\varphi=\pi$, only the $\Phi_3$ spectrum remains. (b)~Eigenstate-resolved rates with the pump tuned to $\Phi_3$ ($E_B \approx 11.6$~meV). At $\varphi = \pi$, $\Phi_3$ reaches its maximum and every bright-state rate vanishes. Away from $\varphi = \pi$, the bright-state rates are weighted by their detunings from the pump.}
\end{figure}

Figure~\ref{fig:bell}(a) shows the total pair-absorption spectrum as a function of pump energy at $\varphi=0$ and $\varphi=\pi$. At $\varphi = 0$ ($\ket{\psi^+}$), the spectrum contains the $\Phi_1$, $\Phi_2$, and $\Phi_6$ contributions with the eigenstate-resolved weights
\begin{equation}
	\label{eq:36}
	W_1(0) = \tfrac{1}{3}\kappa_1,\quad W_2(0) = \tfrac{1}{2}\kappa_2,\quad W_6(0) = \tfrac{1}{6}\kappa_6.
\end{equation}
The prefactors $1/3$, $1/2$, and $1/6$ are the fixed geometric coupling efficiencies $2|g_\alpha|^2$ of the normalized symmetric superposition. They quantify how strongly each eigenstate couples to the cross-circular sector. When the pump targets a specific eigenstate $\Phi_\alpha$, the spectral factor $\kappa_\alpha$ [Eq.~\eqref{eq:28}] is strongly peaked for that eigenstate, and the rate is $W_\alpha(0) = 2|g_\alpha|^2 \kappa_\alpha$. For example, targeting $\Phi_1$ with $\omega_p \approx \omega_{\Phi_1,g}$ gives a rate $W_1(0) = \frac{1}{3}\kappa_1$, where the factor $1/3$ is the intrinsic geometric penalty for coupling to $\Phi_1$.

At $\varphi = \pi$ ($\ket{\psi^-}$), by contrast, every bright eigenstate vanishes individually. For every $\alpha \in \{1,2,6\}$:
\begin{equation}
	\label{eq:37}
	W_\alpha(\pi) = {\kappa_\alpha} \times |\braket{\Phi_\alpha|\psi_\mathrm{prep}(\pi)}|^2 = 0,
\end{equation}
so the targeted bright eigenstate is not populated at $\varphi = \pi$ within the model. Only the $\Phi_3$ spectrum remains at $\varphi=\pi$ in Fig.~\ref{fig:bell}(a). The $\Phi_2$ and $\Phi_3$ lines overlap at the adopted linewidths, but the $\Phi_2$ contribution vanishes at $\varphi=\pi$ by exchange symmetry. This symmetry-based extinction differs in mechanism from spectral selectivity. Detuning the pump frequency away from a bright eigenstate suppresses its rate through the Lorentzian tail of $\kappa_\alpha$, and can achieve excellent contrast when eigenstates are spectrally well resolved relative to their linewidths. The Bell-phase mechanism, by contrast, produces a zero that is enforced by the exchange symmetry and does not rely on the spectral separation of eigenstates.

Figure~\ref{fig:bell}(b) shows the eigenstate-resolved rates as the Bell-state phase is scanned with the pump tuned to $\Phi_3$ ($E_B \approx 11.6$~meV). At $\varphi = \pi$, $W_3(\pi) = \kappa_3$ and every bright-state rate vanishes. At $\varphi = 0$, $\Phi_3$ is suppressed by the symmetry-based extinction. The $\Phi_1$ and $\Phi_6$ rates are also small because these eigenstates are detuned by $\sim 8$~meV from the pump. Only $\Phi_2$ retains an appreciable rate because its 0.7-meV splitting from $\Phi_3$ is comparable to the assumed biexciton linewidth $\hbar\gamma_B \approx 1.5$~meV.

\subsection{\label{sec:classical}Comparison with separable polarization states}

The pumping-rate modulation $W_\alpha(\varphi) \propto 1 \pm \cos\varphi$ for bright and dark cross-circular eigenstates [Eqs.~\eqref{eq:34}--\eqref{eq:35}] originates from the biphoton cross-circular coherence $\rho_{\sigma^+\sigma^-,\sigma^-\sigma^+}$. A separable source with nonzero $\rho_{\sigma^+\sigma^-,\sigma^-\sigma^+}$ can also drive a phase-dependent eigenstate distribution. However, the Peres-Horodecki separability criterion forces any such separable source to carry a compensating cocircular flux, which in turn modifies the eigenstate distribution at the $\Phi_2$, $\Phi_3$, $\Phi_5$, and $\Phi_6$ sectors in ways that the Bell state avoids.

\subsubsection{General separability bound}\label{sec:classical_bound}

A separable polarization state can be decomposed as an incoherent mixture of product states, $\rho_\mathrm{sep} = \sum_k q_k\, \rho_s^{(k)} \otimes \rho_i^{(k)}$ ($q_k \geq 0$, $\sum_k q_k = 1$). Any such state has positive partial transpose (PPT)~\cite{peres1996,horodecki1996}. Applied to the $2\times 2$ principal submatrix of $\rho^{T_i}$ indexed by $\{\ket{\sigma^+\sigma^+}, \ket{\sigma^-\sigma^-}\}$, PPT requires that its determinant be non-negative, giving
\begin{equation}
	\label{eq:38}
	|\rho_{\sigma^+\sigma^-,\sigma^-\sigma^+}|^2 \leq p_{\sigma^+\sigma^+}\,p_{\sigma^-\sigma^-}.
\end{equation}
For two qubits PPT is both necessary and sufficient for separability~\cite{horodecki1996}, so Eq.~\eqref{eq:38} holds for every separable two-photon polarization state, and any violation certifies polarization entanglement of the source. Taking the square root of Eq.~\eqref{eq:38} and applying the inequality $\sqrt{p_{\sigma^+\sigma^+}\,p_{\sigma^-\sigma^-}} \leq (p_{\sigma^+\sigma^+}+p_{\sigma^-\sigma^-})/2$ yields
\begin{equation}
	\label{eq:39}
	2|\rho_{\sigma^+\sigma^-,\sigma^-\sigma^+}| \leq p_{\sigma^+\sigma^+}+p_{\sigma^-\sigma^-} \equiv p_\mathrm{co},
\end{equation}
so any phase-tunable separable source carries cocircular flux of at least twice its cross-circular coherence. The cross-circular Bell state Eq.~\eqref{eq:30} has $|\rho_{\sigma^+\sigma^-,\sigma^-\sigma^+}| = 1/2$ and $p_\mathrm{co} = 0$, violating Eq.~\eqref{eq:39} by the largest possible margin.

\subsubsection{Eigenstate-distribution consequences}\label{sec:classical_eigenstate}

The cocircular analog of Eq.~\eqref{eq:27} follows by summing over the same-helicity pairs $(\sigma^+,\sigma^+) \to \ket{B_1}$ and $(\sigma^-,\sigma^-) \to \ket{B_6}$:
\begin{equation}
	\label{eq:40}
	W_\alpha^{(\mathrm{co})} = \kappa_\alpha^{(\mathrm{co})} \sum_{\substack{(\mu\nu),(\mu'\nu')}} \braket{ \Phi_\alpha | B_{\mu\nu} } \braket{ B_{\mu'\nu'} | \Phi_\alpha } \rho_{\mu\nu,\mu'\nu'},
\end{equation}
where $(\mu\nu),(\mu'\nu') \in \{(\sigma^+\sigma^+),(\sigma^-\sigma^-)\}$ with cocircular populations $p_{\sigma^+\sigma^+}$, $p_{\sigma^-\sigma^-}$ and coherence $\rho_{\sigma^+\sigma^+,\sigma^-\sigma^-}$. We define the cocircular coupling efficiency by $|h_\alpha|^2 \equiv |\braket{\Phi_\alpha | B_1}|^2 = |\braket{\Phi_\alpha | B_6}|^2$, analogous to $|g_\alpha|^2$ in the cross-circular sector.

The Bell state concentrates all excitation on $\Phi_3$ at $\varphi = \pi$, giving the dark-state rate $W_3^{(\mathrm{Bell})}(\varphi = \pi) = \kappa_3^{(\mathrm{cross})}$ (Sec.~\ref{sec:exactfactor}). For any separable source, Eq.~\eqref{eq:35} generalizes to $W_3^{(\mathrm{sep})} = (\kappa_3^{(\mathrm{cross})}/2)\bigl[p_{\sigma^+\sigma^-}+p_{\sigma^-\sigma^+} - 2\mathrm{Re}(\rho_{\sigma^+\sigma^-,\sigma^-\sigma^+})\bigr]$. Using $-\mathrm{Re}(z) \leq |z|$ and Eq.~\eqref{eq:39}, the bracketed term is at most $p_{\sigma^+\sigma^-}+p_{\sigma^-\sigma^+}+p_{\sigma^+\sigma^+}+p_{\sigma^-\sigma^-}=1$. The $\Phi_3$ pumping rate is therefore bounded by
\begin{equation}
	\label{eq:41}
	W_3^{(\mathrm{sep})} \leq \frac{\kappa_3^{(\mathrm{cross})}}{2}, \quad W_3^{(\mathrm{Bell})}(\varphi = \pi) = \kappa_3^{(\mathrm{cross})}.
\end{equation}
Equality is obtained when $\mathrm{Re}(\rho_{\sigma^+\sigma^-,\sigma^-\sigma^+}) = -|\rho_{\sigma^+\sigma^-,\sigma^-\sigma^+}| = -1/4$ and $p_{\sigma^+\sigma^-} = p_{\sigma^-\sigma^+} = p_{\sigma^+\sigma^+} = p_{\sigma^-\sigma^-} = 1/4$. This separable state also extinguishes the bright cross-circular sector. Its cocircular population drives the antisymmetric eigenstate $\Phi_5 = (\ket{B_1} - \ket{B_6})/\sqrt{2}$ because $p_{\sigma^+\sigma^+} + p_{\sigma^-\sigma^-} = 1/2$. Here $|h_5|^2 = 1/2$. The cross-circular overlap is zero. At equal pair flux, the Bell state has a $\Phi_3$ rate twice the separable upper bound and $W_5=0$, whereas a separable state at its maximum $\Phi_3$ rate has $W_5>0$.

To visualize these bounds concretely we take the balanced product state
\begin{equation}
	\label{eq:42}
	\ket{\psi(\theta)}_\mathrm{sep} = \Big(\frac{\ket{\sigma^+}+\ket{\sigma^-}}{\sqrt{2}}\Big)_{s} \otimes \Big(\frac{\ket{\sigma^+}+e^{i\theta}\ket{\sigma^-}}{\sqrt{2}}\Big)_{i}
\end{equation}
as the separable reference, with populations $p_{\sigma^+\sigma^+} = p_{\sigma^+\sigma^-} = p_{\sigma^-\sigma^+} = p_{\sigma^-\sigma^-} = 1/4$ and coherences $\rho_{\sigma^+\sigma^-,\sigma^-\sigma^+} = e^{i\theta}/4$, $\rho_{\sigma^+\sigma^+,\sigma^-\sigma^-} = e^{-i\theta}/4$. Equation~\eqref{eq:42} saturates both Eq.~\eqref{eq:38} [$|\rho_{\sigma^+\sigma^-,\sigma^-\sigma^+}|^2 = 1/16 = p_{\sigma^+\sigma^+}\,p_{\sigma^-\sigma^-}$] and, at $\theta = \pi$, the separability bound Eq.~\eqref{eq:41}. Table~\ref{tab:eigenstate_comparison} compares the eigenstate-resolved rates for the Bell and separable sources.

\begin{figure}[t]
	\includegraphics[width=\columnwidth]{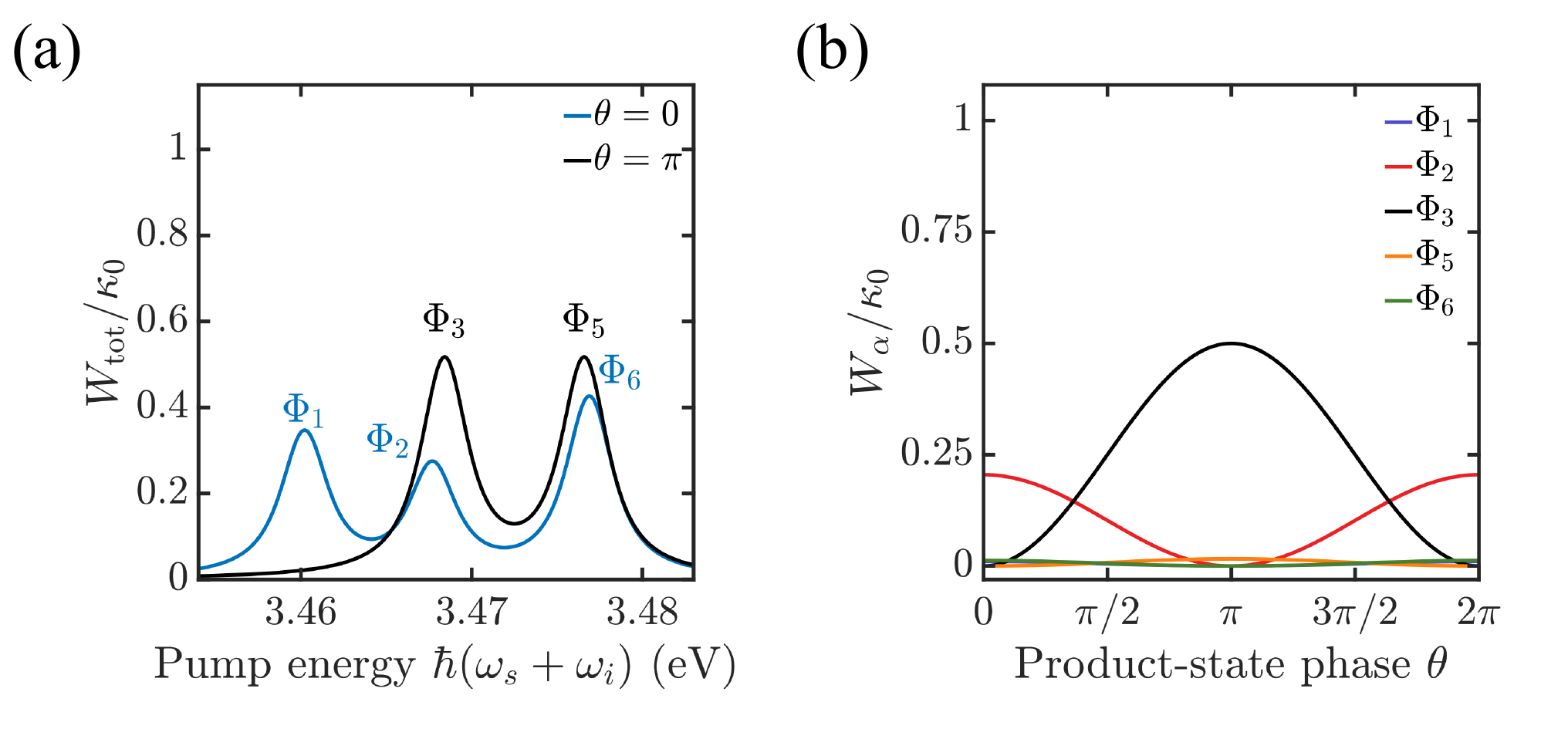}
	\caption{\label{fig:product}%
	Pair-absorption rates for the separable reference state of Eq.~\eqref{eq:42}. (a)~The normalized total pair-absorption rate $W_\mathrm{tot}/\kappa_0$ as a function of pump energy at $\theta=0$ and $\theta=\pi$, using the same linewidths as Fig.~\ref{fig:bell}(a). (b)~Eigenstate-resolved rates with the pump tuned to $\Phi_3$ ($E_B \approx 11.6$~meV).}
\end{figure}

Figure~\ref{fig:product}(a) shows the total pair-absorption spectra of the separable state at $\theta=0$ and $\theta=\pi$. At $\theta=0$, the $\Phi_1$ peak is unchanged from the Bell-state peak at $\varphi=0$ in Fig.~\ref{fig:bell}(a). The $\Phi_2$ peak is reduced by a factor of two. The $\Phi_6$ peak is increased by the cocircular channel. At $\theta=\pi$, the $\Phi_3$ peak reaches half the Bell-state value, consistent with Eq.~\eqref{eq:41}. The cocircular flux also produces a $\Phi_5$ peak. Figure~\ref{fig:product}(b) shows the eigenstate-resolved rates as $\theta$ is scanned with the pump tuned to $\Phi_3$. The $\Phi_3$ rate reaches half the Bell-state maximum at $\theta=\pi$. The nearby $\Phi_2$ rate is largest at $\theta=0$. A smaller $\Phi_5$ contribution remains at $\theta=\pi$, while the $\Phi_1$ and $\Phi_6$ rates are suppressed by spectral detuning.

\begin{table}[htbp]
	\caption{\label{tab:eigenstate_comparison}%
	Eigenstate-resolved pumping rates for the cross-circular Bell state $\ket{\psi(\varphi)}_\mathrm{Bell}$ [Eq.~\eqref{eq:30}] and the separable reference state $\ket{\psi(\theta)}_\mathrm{sep}$ [Eq.~\eqref{eq:42}]. Coupling efficiencies $|g_\alpha|^2$ and $|h_\alpha|^2$ follow from the eigenstates in Table~\ref{tab:eigenstates}.}
	\begin{ruledtabular}
		\begin{tabular}{cll}
			$\Phi_\alpha$
			& \multicolumn{1}{c}{Bell: $W_\alpha^{(\mathrm{Bell})}(\varphi)$}
			& \multicolumn{1}{c}{Separable: $W_\alpha^{(\mathrm{sep})}(\theta)$} \\
			\colrule
			$\Phi_1$
			& $\dfrac{\kappa_1^{(\mathrm{cross})}}{6}(1{+}\cos\varphi)$
			& $(1{+}\cos\theta)\Bigl[\dfrac{\kappa_1^{(\mathrm{cross})}}{12}{+}\dfrac{\kappa_1^{(\mathrm{co})}}{12}\Bigr]$ \\[6pt]
			$\Phi_2$
			& $\dfrac{\kappa_2^{(\mathrm{cross})}}{4}(1{+}\cos\varphi)$
			& $\dfrac{\kappa_2^{(\mathrm{cross})}}{8}(1{+}\cos\theta)$ \\[6pt]
			$\Phi_{3}$
			& $\dfrac{\kappa_3^{(\mathrm{cross})}}{2}(1{-}\cos\varphi)$
			& $\dfrac{\kappa_3^{(\mathrm{cross})}}{4}(1{-}\cos\theta)$ \\[6pt]
			$\Phi_4$
			& $0$ & $0$ \\[6pt]
			$\Phi_5$
			& $0$
			& $\dfrac{\kappa_5^{(\mathrm{co})}}{4}(1{-}\cos\theta)$ \\[6pt]
			$\Phi_6$
			& $\dfrac{\kappa_6^{(\mathrm{cross})}}{12}(1{+}\cos\varphi)$
			& $(1{+}\cos\theta)\Bigl[\dfrac{\kappa_6^{(\mathrm{cross})}}{24}{+}\dfrac{\kappa_6^{(\mathrm{co})}}{6}\Bigr]$ \\
		\end{tabular}
	\end{ruledtabular}
\end{table}

\section{\label{sec:robustness}Decoherence and visibility}

In a real TMD monolayer, the intermediate exciton interacts with its environment before the second photon completes the biexciton transition. The measured signal in any detection channel contains a component set by the eigenstate-selective pumping rate $W_\alpha(\varphi)$~\cite{bamford1986,eichhorn2022,schlawin2013nc,kuhn2020} and a $\varphi$-independent background dominated by one-photon absorption and uncorrelated sequential absorption. We quantify the robustness of the selection rule through the phase-scan visibility of the eigenstate rates,
\begin{equation}
	\label{eq:43}
	V \equiv \frac{\left|W_\alpha(0) - W_\alpha(\pi)\right|}{W_\alpha(0) + W_\alpha(\pi)},
\end{equation}
which measures the depth of the selection-rule extinction. A near-unit value confirms that the exchange-symmetry-based extinction persists under phonon decoherence. Experimental resolvability also depends on the pair-absorption probability and the readout noise.

\subsection{\label{sec:coherence}Entanglement-protected phase transfer}

After the first photon is absorbed from the Bell state, the joint state of (material exciton + remaining idler photon) is
\begin{equation}
	\label{eq:44}
	\begin{split}
		\ket{\psi_{X,i}} = \frac{1}{\sqrt{2}}\big(\ket{X_K} \otimes \ket{\sigma^-}_i + e^{i\varphi} \ket{X_{K'}} \otimes \ket{\sigma^+}_i\big).
	\end{split}
\end{equation}
Because the two terms are correlated with orthogonal idler polarizations, the reduced material state $\rho_\mathrm{mat} = \mathrm{Tr}_i\ket{\psi_{X,i}}\bra{\psi_{X,i}} = (\ket{X_K}\bra{X_K} + \ket{X_{K'}}\bra{X_{K'}})/2$ carries no intervalley coherence; the phase $\varphi$ resides entirely in the material--idler entanglement. When the idler is subsequently absorbed, the valley selection rules [Eq.~\eqref{eq:7}] map $\ket{X_K} + \sigma^- \to \ket{B_2}$ and $\ket{X_{K'}} + \sigma^+ \to \ket{B_3}$, transferring $\varphi$ into the off-diagonal coherence between the two biexciton configurations, $\braket{B_2|\rho|B_3} \propto e^{-i\varphi}$, that determines eigenstate selectivity. By contrast, single-photon valley excitation~\cite{tokman2015} with a linearly polarized photon $(\ket{\sigma^+} + \ket{\sigma^-})/\sqrt{2}$ creates the valley-coherent superposition $(\ket{X_K} + \ket{X_{K'}})/\sqrt{2}$, placing the coherence directly in $\rho_\mathrm{mat}$ where it is exposed to phonon dephasing.

Symmetric exciton damping, the homogeneous dephasing $\gamma_X$ that enters the $K$- and $K'$-valley amplitudes $\mathcal{M}_2$ and $\mathcal{M}_3$ identically through Eq.~\eqref{eq:26}, reduces the overall excitation rate but preserves the ratio $\mathcal{M}_3/\mathcal{M}_2 = e^{i\varphi}$ by valley symmetry and therefore does not degrade the eigenstate selectivity.

Within the model considered here, the primary mechanism that degrades the visibility is which-valley information leakage. If the phonon environment, initially in a common state $\ket{\mathcal{E}_0}$, evolves into valley-dependent states $\ket{\mathcal{E}_K}$ and $\ket{\mathcal{E}_{K'}}$,
\begin{equation}
	\label{eq:45}
	\begin{aligned}
		\ket{X_K}\ket{\mathcal{E}_0} &\to \ket{X_K}\ket{\mathcal{E}_K}, \\
		\ket{X_{K'}}\ket{\mathcal{E}_0} &\to \ket{X_{K'}}\ket{\mathcal{E}_{K'}},
	\end{aligned}
\end{equation}
the biexciton off-diagonal coherence acquires a suppression factor $\braket{\mathcal{E}_K|\mathcal{E}_{K'}}$, which sets the phase-scan visibility (Appendix~\ref{sec:appD1}). The extent of this leakage depends on how long the intermediate exciton is exposed to the phonon bath before the second photon arrives. The sinc-shaped JSA of type-II SPDC has a total temporal correlation window $[-T_e/2,\, T_e/2]$; causality (the second photon must arrive after the first) restricts the intermediate-state dwell time to the positive half $[0,\, T_e/2]$ (Appendix~\ref{sec:appB}). The exciton dephasing further down-weights long dwell times. The resulting mean intermediate-exciton dwell time (Appendix~\ref{sec:appD1}) is
\begin{equation}
	\label{eq:46}
	\langle\tau\rangle = \frac{1}{\gamma_X} - \frac{T_e/2}{e^{\gamma_X T_e/2} - 1}.
\end{equation}
For broadband SPDC ($T_e \sim 100$ fs) with $\hbar\gamma_X \sim 1.5$ meV, $\gamma_X T_e \approx 0.23 \ll 1$, placing this regime in the broad-bandwidth limit $\langle\tau\rangle \to T_e/4$ of Eq.~\eqref{eq:46}, giving $\langle\tau\rangle \approx 25$~fs, one to two orders of magnitude shorter than picosecond valley-coherence times, so the intermediate exciton has minimal exposure to valley decoherence. In the narrow-bandwidth limit, $\langle\tau\rangle$ instead saturates at $1/\gamma_X \approx 0.44$~ps. This exceeds the shortest reported valley-coherence time $\tau_{vc}=0.1$~ps~\cite{hao2016}, so the dwell-time protection is lost in that case. The broadband pair avoids this. The idler arrives within $T_e$ of the signal, which keeps $\langle\tau\rangle \approx 25$~fs at resonance.

\subsection{\label{sec:decoherence}Valley dephasing and intervalley scattering}

Valley dephasing and intervalley scattering both reduce the material-side visibility. Valley dephasing accumulates which-valley information at rate $\gamma_\mathrm{valley}$ during $\langle\tau\rangle$. The valley depolarization time $\tau_v$ measured by Kerr rotation ($\tau_v \approx 6$ ps for the bright $A$-exciton in WSe$_2$ at 4 K~\cite{zhu2014}) characterizes population relaxation between valleys, driven primarily by the Coulomb exchange (Maialle--Silva--Sham) mechanism~\cite{glazov2014,maialle1993}. The valley coherence time $\tau_{vc}$, which governs the decay of intervalley superpositions, is generally shorter, with $1/\tau_{vc} = 1/(2\tau_v) + \gamma_v^*$, where $\gamma_v^*$ is the pure valley dephasing rate~\cite{glazov2014,wang2018}. The quantity entering the model is the valley decoherence rate $\gamma_\mathrm{valley} \equiv 1/\tau_{vc}$, i.e., the rate at which the phonon environment acquires which-valley information. Averaging the exponential decoherence factor over the assumed dwell-time distribution gives (Appendix~\ref{sec:appD1})
\begin{equation}
	\label{eq:47}
	V_\mathrm{deph} = \frac{\gamma_X}{\gamma_X + \gamma_\mathrm{valley}} \frac{1 - e^{-(\gamma_X + \gamma_\mathrm{valley})T_e/2}} {1 - e^{-\gamma_X T_e/2}}.
\end{equation}
With $\hbar\gamma_X \approx 1.5$ meV for WSe$_2$~\cite{moody2015}, we explore $\tau_{vc} = 0.1$--$3$ ps ($\hbar\gamma_\mathrm{valley} \approx 0.2$--$6.6$ meV) for the valley-coherence time, covering the direct two-dimensional coherent-spectroscopy value $\tau_{vc} \approx 0.1$ ps~\cite{hao2016} and the picosecond-scale values inferred from photoluminescence linear polarization in high-quality samples~\cite{jones2013}.

Intervalley scattering physically transfers an exciton between valleys, corrupting the pathway mapping. In monolayer WSe$_2$, this requires large momentum transfer and is mediated by short-range disorder or zone-edge phonons, placing the intervalley scattering time $\tau_\mathrm{iv}=1/\gamma_\mathrm{iv}$ at or above the measured valley depolarization time $\tau_v \approx 6$ ps at 4 K~\cite{zhu2014}. The scattering probability follows by averaging $e^{-\gamma_\mathrm{iv}\tau}$ over the intermediate-state dwell-time distribution (see Appendix~\ref{sec:appD2})
\begin{equation}
	\label{eq:48}
	P_\mathrm{iv} = 1 - \frac{\gamma_X}{\gamma_X + \gamma_\mathrm{iv}} \frac{1 - e^{-(\gamma_X + \gamma_\mathrm{iv})T_e/2}} {1 - e^{-\gamma_X T_e/2}},
\end{equation}
giving $P_\mathrm{iv} \approx 0.2\%$ for broadband SPDC when $\tau_\mathrm{iv} = 10$~ps is used. If $\gamma_\mathrm{iv}$ represents a pathway-changing channel independent of the processes already contained in the chosen $\tau_{vc}$, the combined material-side visibility is
\begin{equation}
	\label{eq:49}
	V_\mathrm{mat} = V_\mathrm{deph} \times (1 - P_\mathrm{iv}).
\end{equation}
The model rate for an ideal source is then $W_\alpha(\varphi) \propto 1 \pm V_\mathrm{mat}\cos\varphi$, with the upper sign for the bright sector and the lower sign for the dark eigenstate.

\begin{figure}[t]
	\includegraphics[width=\columnwidth]{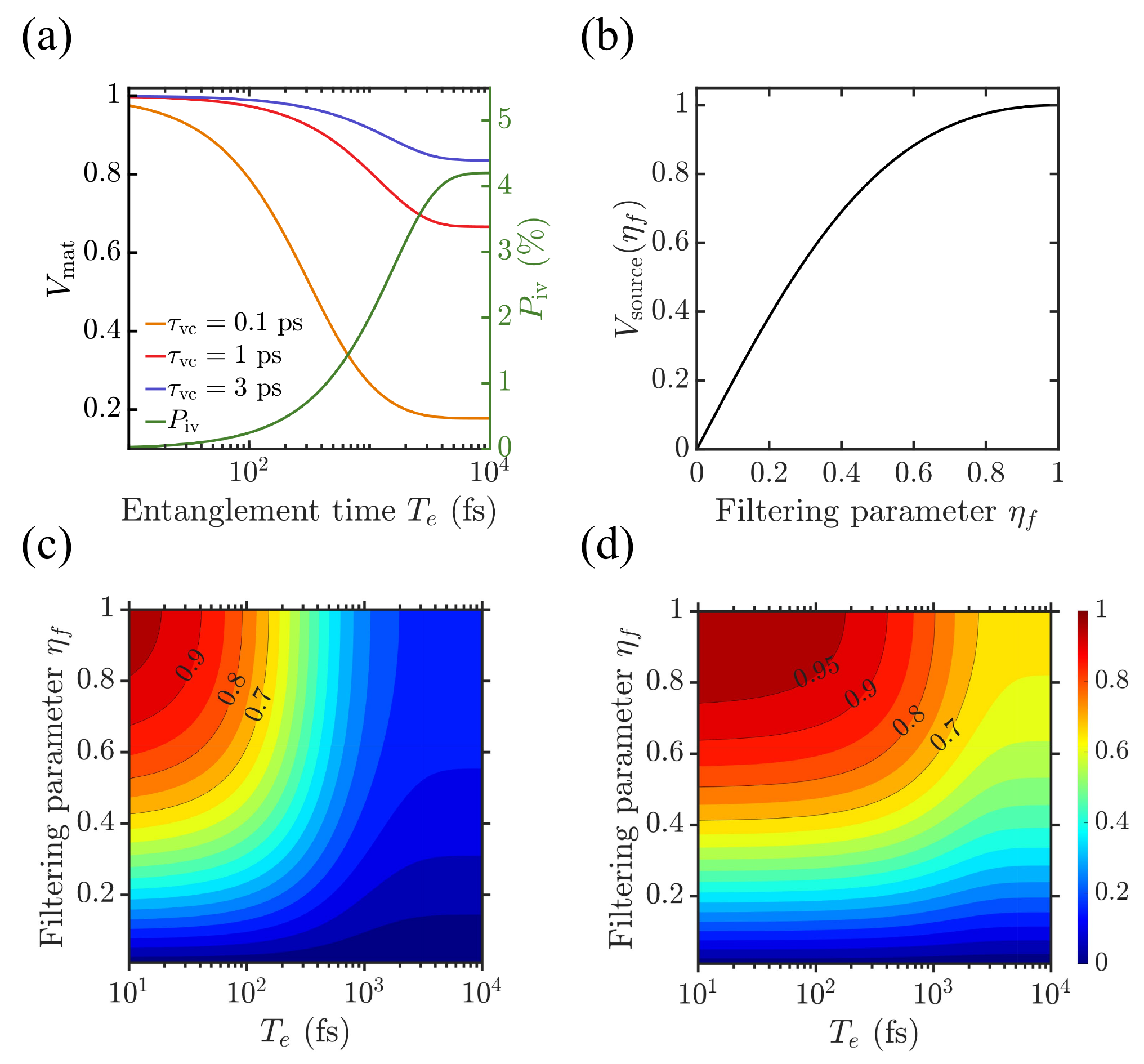}
	\caption{\label{fig:visibility}%
	Decoherence and source-imperfection limits on the eigenstate selection rule. (a)~Material-side visibility $V_\mathrm{mat}$ (left axis) as a function of entanglement time $T_e$ for valley coherence times $\tau_{vc} = 0.1$~ps~\cite{hao2016} and $1$ and $3$~ps~\cite{jones2013} at 4~K. The intervalley scattering probability $P_\mathrm{iv}$ (right axis, green) remains below $\sim 4\%$ across the full range. (b)~Source-side visibility $V_\mathrm{source}(\eta_f)$ as a function of the polarization-filtering parameter $\eta_f$. (c),~(d)~The model visibility $V_\mathrm{model}=V_\mathrm{mat}V_\mathrm{source}V_\mathrm{res}$ as a function of both $T_e$ and $\eta_f$ for $\tau_{vc} = 0.1$~ps~(c) and $1$~ps~(d). The residual-population factor is $V_\mathrm{res}=0.9956$. Contour lines mark $V_\mathrm{model} = 0.70$, $0.80$, $0.90$, and $0.95$.}
\end{figure}

The material-side visibility $V_\mathrm{mat}$ and $P_\mathrm{iv}$ vary with $T_e$ [Fig.~\ref{fig:visibility}(a)]. In the broadband regime ($T_e \lesssim 100$~fs), the $\tau_{vc} = 1$ and $3$~ps curves converge near $V_\mathrm{mat} \approx 0.97\text{--}0.99$ because the intermediate-exciton dwell time is so short ($\langle\tau\rangle \lesssim 25$~fs) that valley decoherence barely acts, while the $\tau_{vc} = 0.1$~ps curve is already reduced to $V_\mathrm{mat} \approx 0.79$. As $T_e$ increases, the curves separate because the exciton is exposed to the phonon bath for longer dwell times. The curve for $\tau_{vc}=0.1$~ps decreases most rapidly, followed by those for $1$ and $3$~ps. In the long-entanglement-time limit ($T_e\to\infty$), $V_\mathrm{deph}$ approaches approximately $0.19$, $0.70$, and $0.87$ for $\tau_{vc}=0.1$, $1$, and $3$~ps, respectively. Including intervalley scattering, the corresponding limits of $V_\mathrm{mat}$ are approximately $0.18$, $0.67$, and $0.84$. The intervalley scattering probability $P_\mathrm{iv}$ is independent of $\tau_{vc}$ and remains below $\sim 4\%$ across the entire $T_e$ range.

\subsection{\label{sec:relaxation}Dark-to-bright relaxation within the triplet pair}

Under triplet-tuned excitation, the relevant relaxation channel for the dark eigenstate $\Phi_3$ is its exchange-split partner $\Phi_2$, which lies only about $0.7$~meV below $\Phi_3$ as shown in Table~\ref{tab:eigenstates}. Because the splitting is only $0.7$~meV, the $\Phi_3 \to \Phi_2$ transfer is not suppressed by an energy gap, and its rate is a principal kinetic unknown of the scheme. The exchange model fixes the composition and splittings of the eigenstates but does not determine this rate; reported biexciton lifetimes in monolayer TMDs lie in the picosecond regime~\cite{you2015,wang2018,he2016}, yet the lifetime of the exchange-dark eigenstate $\Phi_3$ has not been measured directly.

The transfer occurs after the absorption event and does not change the pumping rates $W_\alpha(\varphi)$. It reduces the $\Phi_3$ population and produces a secondary $\Phi_2$ population. This secondary population follows the phase dependence of the prepared $\Phi_3$ population. It is distinct from the $\Phi_2$ population created in the pair-absorption event. We denote the transfer rate from $\Phi_3$ to $\Phi_2$ by $\Gamma_{3\to2}$. Slow transfer satisfies
\begin{equation}
	\label{eq:50}
	\Gamma_{3\to2} \ll \Gamma_\mathrm{loss}^{(\Phi_3)},
\end{equation}
where $\Gamma_\mathrm{loss}^{(\Phi_3)}$ denotes the total rate of nonradiative decay and leakage out of the biexciton manifold. When this inequality is satisfied, transfer is a subdominant contribution to the decay of $\Phi_3$. At $\varphi=\pi$, the pair-absorption rate into $\Phi_2$ vanishes [Eq.~\eqref{eq:37}]. A time-resolved measurement of the subsequent $\Phi_2$ population can therefore determine $\Gamma_{3\to2}$.

\subsection{\label{sec:source}Polarization-dependent loss and partial entanglement}

Polarization-dependent losses in the excitation path (stress birefringence in optical elements, imperfect wave-plate alignment, or unbalanced mirror reflectivities) attenuate one circular component relative to the other. We model this as a local filter $F = \mathrm{diag}(1,\eta_f)$ acting on the signal mode ($0 < \eta_f \le 1$), which transforms the Bell state $\ket{\psi(\varphi)}_\mathrm{Bell}$ [Eq.~\eqref{eq:30}] into
\begin{equation}
	\label{eq:51}
	\ket{\psi'} \propto \ket{\sigma^+}_s\ket{\sigma^-}_i + \eta_f\, e^{i\varphi}\ket{\sigma^-}_s\ket{\sigma^+}_i,
\end{equation}
since $F$ passes $\ket{\sigma^+}$ unchanged and attenuates $\ket{\sigma^-}$ by $\eta_f$. After normalization, the polarization density matrix elements become
\begin{equation}
	\label{eq:52}
	\begin{aligned}
		p_{\sigma^+\sigma^-} = \frac{1}{1+\eta_f^2}&, \quad p_{\sigma^-\sigma^+} = \frac{\eta_f^2}{1+\eta_f^2}, \\
		\rho_{\sigma^+\sigma^-,\sigma^-\sigma^+}
		&= \frac{\eta_f\, e^{-i\varphi}}{1+\eta_f^2}.
	\end{aligned}
\end{equation}
Substituting into Eq.~\eqref{eq:32} and using the exchange symmetry $\braket{\Phi_\alpha | B_2} = \braket{\Phi_\alpha | B_3} \equiv g_\alpha$ for the bright eigenstates gives
\begin{equation}
	\label{eq:53}
	\begin{split}
		W_\alpha(\varphi) &= \kappa_\alpha |g_\alpha|^2 \left[p_{\sigma^+\sigma^-} + p_{\sigma^-\sigma^+} + 2\mathrm{Re}\left(\rho_{\sigma^+\sigma^-,\sigma^-\sigma^+}\right) \right] \\
		&= \kappa_\alpha |g_\alpha|^2 \left[1 + \frac{2\eta_f}{1+\eta_f^2}\cos\varphi\right].
	\end{split}
\end{equation}
Reading off the coefficient of $\cos\varphi$ yields the visibility
\begin{equation}
	\label{eq:54}
	V_\mathrm{source}(\eta_f) = \frac{2\eta_f}{1+\eta_f^2} = \mathrm{sech}(\gamma), \quad \eta_f = e^{-\gamma},
\end{equation}
which interpolates smoothly from unity (perfect source, $\eta_f = 1$) to zero (complete attenuation of one circular component, $\eta_f \to 0$), as shown in Fig.~\ref{fig:visibility}(b). The concurrence $C$ is a standard measure of two-qubit entanglement, ranging from 0 for a separable state to 1 for a maximally entangled Bell state~\cite{wootters1998}. Under the local filtering $F = \mathrm{diag}(1,\eta_f)$, the concurrence of the filtered state is $C'(\eta_f) = 2\eta_f/(1+\eta_f^2)$~\cite{verstraete2001}. It is equal to $V_\mathrm{source}(\eta_f)$ in Eq.~\eqref{eq:54}. This equality applies to an initial pure Bell state under local filtering on one arm. Within the local-filtering model used here, a reduction in concurrence therefore gives the same reduction in the eigenstate-selection visibility.

\subsection{\label{sec:budget}Residual bright population and model visibility}

Ideally, the bright-eigenstate population vanishes at $\varphi=\pi$. Unequal valley responses and wave-plate errors leave a nonzero bright-sector population at this phase. We refer to this leakage as the residual bright population. In the sample, a valley splitting or a difference in the valley dipoles or dephasing rates makes the two diagonal valley responses unequal, $T_2\neq T_3$. In the optics, a wave-plate axis misalignment $\theta_\mathrm{err}$ or a retardance deviation $\delta$ mixes a small cocircular amplitude into the state arriving at the sample. Defining the difference between the two-photon valley kernels as $\delta T(\omega)\equiv T_2(\omega)-T_3(\omega)$, the bright amplitude at $\varphi=\pi$ [Eqs.~\eqref{eq:16a} and~\eqref{eq:16b}] reduces to
\begin{equation}
	\label{eq:55}
	\mathcal{M}_2+\mathcal{M}_3\big|_{\varphi=\pi}=\frac{1}{\sqrt{2}}\iint d\omega_s\,d\omega_i\; f\,\big[\delta T(\omega_s)-\delta T(\omega_i)\big]\,L.
\end{equation}
This sum is linear in $\delta T$, so the residual bright population scales as $|\delta T/T|^2$. The wave-plate contributions are likewise quadratic and give $(2\theta_\mathrm{err})^2$ and $(\delta/2)^2$. Appendix~\ref{sec:appE} gives the numerical estimates.

We quantify this leakage by the normalized residual fraction $P_\mathrm{res}\equiv W_\mathrm{bright}(\pi)/[W_\mathrm{bright}(0)+W_\mathrm{bright}(\pi)]$. Table~\ref{tab:budget} gives $P_\mathrm{res}\approx2.2\times10^{-3}$ from the three amplitude errors. The amplitude errors alone leave a dark-sector population $1-P_\mathrm{res}\approx0.998$ at $\varphi=\pi$.

Polarization-dependent loss and spectral distortion reduce the coherence between the cross-circular pathways. Writing $|\rho_{\sigma^+\sigma^-,\sigma^-\sigma^+}|=(1-\epsilon)/2$, Eq.~\eqref{eq:32} gives
\begin{equation*}
	W_\mathrm{bright}(\varphi)\propto1+(1-\epsilon)\cos\varphi.
\end{equation*}
The source-side visibility is therefore $V_\mathrm{source}=1-\epsilon$. Spectral distinguishability enters the same density-matrix element. Both effects can be included through the full polarization density matrix measured at the sample.

\begin{table*}[htbp]
	\caption{\label{tab:budget}%
	Error budget at $\varphi=\pi$. Amplitude errors enter quadratically through $P_\mathrm{res}$. Coherence loss enters linearly through $1-V_\mathrm{source}$.}
	\begin{ruledtabular}
		\begin{tabular}{lll}
			Imperfection & Input value & Contribution \\
			\hline
			\multicolumn{3}{l}{\textit{Amplitude errors (quadratic)}} \\
			Valley-response difference $\delta T/T$ & $|\delta T/T|=10^{-2}$ & $|\delta T/T|^2=10^{-4}$ \\
			Wave-plate angle error & $\theta_\mathrm{err}=1^\circ$ & $(2\theta_\mathrm{err})^2\approx1.2\times10^{-3}$ \\
			Wave-plate retardance error & $\delta\approx0.06$~rad ($\lambda/100$) & $(\delta/2)^2\approx9\times10^{-4}$ \\
			\multicolumn{3}{l}{\textit{Coherence loss (linear)}} \\
			Two-photon polarization coherence at the sample & $V_\mathrm{source}$ from the polarization density matrix & $1-V_\mathrm{source}$ \\
\hline
Total amplitude errors & & $P_\mathrm{res}\approx2.2\times10^{-3}$ \\
Dark-sector population from amplitude errors $1-P_\mathrm{res}$ & & $\approx0.998$ \\
		\end{tabular}
	\end{ruledtabular}
\end{table*}

Appendix~\ref{sec:appE} derives $V_\mathrm{res}=1-2P_\mathrm{res}$ from the bright-sector rates at $\varphi=0$ and $\varphi=\pi$. The model visibility is
\begin{equation}
	\label{eq:56}
	\begin{split}
		V_\mathrm{model}
		&= V_\mathrm{mat}V_\mathrm{source}(\eta_f)V_\mathrm{res} \\
		&= V_\mathrm{mat}V_\mathrm{source}(\eta_f)(1-2P_\mathrm{res}).
	\end{split}
\end{equation}
Figures~\ref{fig:visibility}(c) and~\ref{fig:visibility}(d) show $V_\mathrm{model}$ after the residual amplitude errors are included. For $\tau_{vc}=0.1$~ps and $\eta_f\gtrsim0.95$, $V_\mathrm{model}\gtrsim0.90$ requires $T_e\lesssim40$~fs. At $T_e\sim100$~fs and $\eta_f\gtrsim0.95$, the model gives $V_\mathrm{model}\approx0.79$ for $\tau_{vc}=0.1$~ps and $V_\mathrm{model}\gtrsim0.956$ for $\tau_{vc}=1$~ps.

\section{\label{sec:experimental}Experimental considerations}

To our knowledge, no experimental value of the two-photon absorption cross section has been reported for monolayer WSe$_2$. To estimate the signal level without this value, we define the pair-absorption probability
\begin{equation*}
	P_\mathrm{abs}(\varphi)\equiv\frac{W_\mathrm{tot}(\varphi)}{R_\mathrm{pairs}},\qquad W_\mathrm{tot}(\varphi)=\sum_\alpha W_\alpha(\varphi),
\end{equation*}
where $W_\mathrm{tot}(\varphi)$ is the total absorption rate and $R_\mathrm{pairs}$ is the incident SPDC pair rate.

The coincidence readout counts the transmitted signal--idler pairs in coincidence [Fig.~\ref{fig:setup}(b)]. Pair absorption removes both photons of an absorbed pair, so the coincidence rate drops by the fractional depth $P_\mathrm{abs}(\varphi)$. The pair-absorption contribution to the dip therefore depends on the total rate $W_\mathrm{tot}$. The bright-state rates vary as $(1+\cos\varphi)/2$, while the $\Phi_3$ rate varies as $(1-\cos\varphi)/2$. Their sum is
\[
	\begin{aligned}
		W_\mathrm{tot}(\varphi)&=\tfrac{1}{2}(\kappa_\mathrm{bright}+\kappa_3)+\tfrac{1}{2}(\kappa_\mathrm{bright}-\kappa_3)\cos\varphi,\\
		\kappa_\mathrm{bright}&\equiv2\!\!\sum_{\alpha\in\{1,2,6\}}\!\!\kappa_\alpha|g_\alpha|^2.
	\end{aligned}
\]
With the pump tuned to $\Phi_3$, the Lorentzian factors of Fig.~\ref{fig:bell}(a) at $\hbar\gamma_B=1.5$~meV give $\kappa_\mathrm{bright}\approx0.43\,\kappa_3$, dominated by the nearby $\Phi_2$ line. The total rate therefore retains a $\varphi$ dependence.

The coincidence dip also contains $\varphi$-independent contributions from single-photon absorption of the resonant signal and uncorrelated sequential absorption. To remove their mean contributions, we use the differential signal $\Delta P_\mathrm{abs}=|P_\mathrm{abs}(\pi)-P_\mathrm{abs}(0)|$. With true and accidental coincidence counts $N_c$ and $N_a$ at each phase, the Poisson-limited detection condition is (Appendix~\ref{sec:appF})
\begin{equation}
	\label{eq:57}
	\mathrm{SNR}_\mathrm{coinc}=\frac{\Delta P_\mathrm{abs}N_c}{\sqrt{2(N_c+N_a)}}\gtrsim1.
\end{equation}
Using a pair rate $R_\mathrm{pairs}=10^{10}~\mathrm{s}^{-1}$, a per-photon detection efficiency $\eta_\mathrm{det}=0.1$, a coincidence window $\tau_w=100$~ps, and an integration time $T_\mathrm{int}=1$~s (Table~\ref{tab:parameters}), we obtain $P_\mathrm{abs}(\pi)\gtrsim3.5\times10^{-4}$. Appendix~\ref{sec:appF} gives the count-rate calculation and estimates a $\varphi$-independent single-photon absorption probability of approximately $10^{-1}$ at the $A$-exciton resonance. Although the single-photon absorption probability is much larger than the detection threshold, its mean contribution cancels in $\Delta P_\mathrm{abs}$.

The coincidence readout measures $\Delta P_\mathrm{abs}$ but does not determine the $\Phi_3$ population that remains after nonradiative decay or transfer to $\Phi_2$. The pump-probe readout complements this measurement by probing the $\Phi_3$ population [Fig.~\ref{fig:setup}(c)]. SPDC photon pairs act as the pump. The observable is the fractional change in the transmission of a weak probe near the $A$-exciton resonance, $\Delta T/T=(T_\mathrm{on}-T_\mathrm{off})/T_\mathrm{off}$. Here $T_\mathrm{on}$ and $T_\mathrm{off}$ are the probe transmissions with the SPDC excitation turned on and off. Although $\ket{\Phi_3}$ is dark in emission, it contains valley excitons and changes the probe transmission through phase-space filling. The steady-state population of $\ket{\Phi_3}$ is
\[
	N_3=\frac{W_3}{\Gamma_\mathrm{loss}^{(\Phi_3)}+\Gamma_{3\to2}},
\]
where $\Gamma_\mathrm{loss}^{(\Phi_3)}$ is the total rate of nonradiative decay and leakage out of the biexciton manifold and $\Gamma_{3\to2}$ is the transfer rate from $\Phi_3$ to $\Phi_2$. The pair-induced part of the probe response is proportional to this population, $\Delta T/T\propto N_3(\varphi)$.

The pump-probe readout has no accidental-coincidence background. It resolves the phase dependence when (Appendix~\ref{sec:appF})
\begin{equation}
	\label{eq:58}
	\left|\left(\frac{\Delta T}{T}\right)_{\!\pi}-\left(\frac{\Delta T}{T}\right)_{\!0}\right|\gtrsim\sigma_\mathrm{pp}.
\end{equation}
For $R_\mathrm{probe}\sim10^{15}~\mathrm{s}^{-1}$ and one second of integration, the probe shot-noise floor is $\sigma_\mathrm{pp}\approx6.3\times10^{-8}$. Laser-intensity noise and detector-response drift can raise this floor.

Bright biexcitons created in the pair-absorption event can also contribute to $\Delta T/T$. Separating their contribution from the $\Phi_3$ response requires measurements of the bright- and dark-biexciton lifetimes and $\Gamma_{3\to2}$.

\begin{table}[htbp]
	\caption{\label{tab:parameters}%
	Parameters used for the signal and visibility estimates.}
	\begin{ruledtabular}
		\begin{tabular}{@{}p{0.53\columnwidth}p{0.41\columnwidth}@{}}
			Parameter & Value \\
			\hline
			\multicolumn{2}{l}{\textit{SPDC source}} \\
			Entanglement time $T_e$ & $100$~fs \\
			SPDC pair rate $R_\mathrm{pairs}$ & $10^{10}~\mathrm{s}^{-1}$ \\
			Coincidence window $\tau_w$ & $100$~ps \\
			Integration time $T_\mathrm{int}$ & $1$~s \\
			Detection efficiency $\eta_\mathrm{det}$ & $0.1$ \\
			Source concurrence $C$ & $\approx0.993$~\cite{fedrizzi2007} \\
			Wave-plate angle error $\theta_\mathrm{err}$ & $1^\circ$ \\
			Retardance tolerance & $\lambda/100$ \\
			\multicolumn{2}{l}{\textit{WSe$_2$ sample}} \\
			Exciton linewidth $\hbar\gamma_X$ & $\approx1.5$~meV~\cite{moody2015} \\
			Exciton lifetime $\tau_X$ & $\sim2$~ps~\cite{robert2016} \\
			Biexciton linewidth $\hbar\gamma_B$ & $\approx1.5$~meV ($=\hbar\gamma_X$) \\
			Gaussian broadening FWHM $\Sigma$ & $1$~meV~\cite{ajayi2017} \\
			Valley coherence \ensuremath{\tau_{vc}} & $0.1$--$3$~ps~\cite{hao2016,jones2013} \\
			Intervalley scattering time $\tau_\mathrm{iv}$ & $10$~ps \\
			Exciton $g$-factor $\lvert g\rvert$ & $\approx4$~\cite{srivastava2015,aivazian2015} \\
			\multicolumn{2}{l}{\textit{Pump--probe readout}} \\
			Probe rate $R_\mathrm{probe}$ & $10^{15}~\mathrm{s}^{-1}$ \\
		\end{tabular}
	\end{ruledtabular}
\end{table}

\section{\label{sec:discussion}Discussion and Conclusion}

We have shown that the Bell-state phase selects the bright and exchange-dark biexciton sectors in monolayer WSe$_2$. At equal pair flux, the $\Phi_3$ pumping rate of the Bell state is twice the upper bound for separable polarization states [Eq.~\eqref{eq:41}]. The estimates in Secs.~\ref{sec:robustness} and~\ref{sec:experimental} give the expected visibility and signal levels for realistic source and sample parameters.

The ongoing effort to measure ETPA cross sections in molecular systems~\cite{landes2021,parzuchowski2021,he2024,mikhaylov2022} targets a quantitative rate enhancement that must be separated from single-photon backgrounds. The eigenstate selection rule proposed here offers a structurally different approach. It uses the Bell-state phase to redistribute excitation between biexciton eigenstates with different exchange symmetry. The resulting phase-dependent change in the biexciton spectrum is a spectroscopic signature. The unknown pair-absorption probability remains an experimental limitation because the spectral change must exceed the readout noise.

The numerical binding energies and exchange splittings in Table~\ref{tab:eigenstates} are computed for WSe$_2$ on sapphire~\cite{steinhoff2018}. Whether each eigenstate is symmetric or antisymmetric under $B_2 \leftrightarrow B_3$ is determined to leading order by the crystal symmetry of the TMD monolayer rather than the substrate. The bright-state extinction at $\varphi=\pi$ and the dark-state extinction at $\varphi=0$ are therefore insensitive to the substrate. Only the quantitative binding energies and visibility estimates should shift under hBN encapsulation. The exciton linewidth used in this work, $\hbar\gamma_X \approx 1.5$~meV, is chosen close to the residual homogeneous linewidth $1.6\pm0.3$~meV measured in WSe$_2$ grown by chemical vapor deposition on sapphire~\cite{moody2015}. The biexciton linewidth is assumed equal to this value. The selection rule relies on valley symmetry, $T_2(\omega) = T_3(\omega)$. Nonmagnetic strain and dielectric disorder that shift both valley responses together broaden the spectrum without lifting the bright-state extinction at $\varphi=\pi$ or the dark-state extinction at $\varphi=0$. Magnetic fields, magnetic disorder, and photoinduced valley polarization can make the two valley responses unequal, $\delta T\neq0$. The asymmetry $\delta T/T$ induced by a stray field below $1$~mT is estimated in Appendix~\ref{sec:appE} at ${\approx}10^{-4}$.

Finally, the symmetry argument is not unique to WSe$_2$. The required ingredients are (i) two degenerate absorption pathways that create the same final state in different internal configurations, (ii) an exchange interaction that splits the configuration space into symmetric and antisymmetric eigenstates, and (iii) a pre-characterized photon source whose relative phase controls those pathways. Other TMD monolayers (MoSe$_2$, WS$_2$, MoS$_2$)~\cite{you2015,hao2017,barbone2018,chen2018,erkensten2021,shahnazaryan2017} share the same valley-based symmetry logic and are natural candidates.

\begin{acknowledgments}
	This work was supported by the Institute of Information \& Communications Technology Planning \& Evaluation (IITP) under Grant No. RS-2025-25464869, funded by the Korea government (Ministry of Science and ICT).
\end{acknowledgments}

\section*{Data Availability}

The data that support the findings of this article are not publicly available. The data are available from the authors upon reasonable request.

\appendix

\section{\label{sec:appA}Two-Photon Transition Amplitude}

Inserting a complete set of intermediate one-exciton states $\sum_n \ket{n}\bra{n} = \mathbb{I}$ between the two interactions in Eq.~\eqref{eq:5}, the interaction-picture time dependence generates transition-frequency phases:
\begin{equation}
	\label{eq:A1}
	\begin{split}
		&\bra{B_j}\hat{d}_{\sigma_2}(t_2)\hat{d}_{\sigma_1}(t_1)\ket{g} \\
		&\quad= \sum_n \bra{B_j}\hat{d}_{\sigma_2}\ket{n}\bra{n}\hat{d}_{\sigma_1}\ket{g} e^{i\omega_{B_j n}t_2} e^{i\omega_{ng}t_1},
	\end{split}
\end{equation}
where $\omega_{B_j n} = (E_{B_j} - E_n)/\hbar$ and $\omega_{ng} = (E_n - E_g)/\hbar$. After the first interaction at $t_1$, the system spends time $\tau = t_2 - t_1$ in the intermediate state $\ket{n}$, during which the optical coherence $\rho_{ng}$ dephases at rate $\gamma_n$. After the second interaction at $t_2$, the system spends time $\tau' = t - t_2$ in the final state $\ket{B_j}$, during which $\rho_{B_j g}$ dephases at rate $\gamma_B$. Including these causal decay factors:
\begin{equation}
	\label{eq:A2}
	\begin{split}
		e^{i\omega_{B_j n}t_2} e^{i\omega_{ng}t_1} &\to e^{i\omega_{B_j n}t_2} e^{i\omega_{ng}t_1} e^{-\gamma_n(t_2 - t_1)} e^{-\gamma_B(t - t_2)}.
	\end{split}
\end{equation}
In the interaction picture, each positive-frequency field operator carries a phase $e^{-i\omega t}$ at the time of the interaction, so the signal photon annihilated at $t_2$ contributes $e^{-i\omega_s t_2}$ and the idler photon annihilated at $t_1$ contributes $e^{-i\omega_i t_1}$. Combining these field phases with the matrix element Eq.~\eqref{eq:10} for Contraction A ($\hat{a}_s$ at $t_2$, $\hat{a}_i$ at $t_1$), the matter response Eq.~\eqref{eq:A2}, and the $(-i/\hbar)^2$ prefactor from Eq.~\eqref{eq:5}:
\begin{equation}
	\label{eq:A3}
	\begin{split}
		\mathcal{M}_j^{(A)}(t) &= \frac{-1}{\hbar^2} \sum_n \iint d\omega_s\, d\omega_i\; \Psi(\omega_s, \sigma_2; \omega_i, \sigma_1) \\
		&\quad\times \bra{B_j}\hat{d}_{\sigma_2}\ket{n} \bra{n}\hat{d}_{\sigma_1}\ket{g} \int_{-\infty}^{t} dt_2 \int_{-\infty}^{t_2} dt_1 \\
		&\quad\times e^{-\gamma_B(t-t_2)} e^{i\omega_{B_j n}t_2} e^{i\omega_{ng}t_1} \\
		&\quad\times e^{-\gamma_n(t_2-t_1)} e^{-i\omega_s t_2} e^{-i\omega_i t_1}.
	\end{split}
\end{equation}

Switching to the dwell-time variables $\tau$ and $\tau'$ introduced above as integration variables, \[ t_2 = t - \tau', \quad t_1 = t - \tau' - \tau, \quad \tau,\tau' \geq 0, \] the Jacobian is unity and both integration limits become $[0, \infty)$.

The final-state damping ($t_2 \to t$) becomes \[ e^{-\gamma_B(t-t_2)} = e^{-\gamma_B\tau'}. \] The intermediate-state phases give
\begin{align*}
	e^{i\omega_{B_j n}t_2} e^{i\omega_{ng}t_1} &= e^{i(\omega_{B_j n}+\omega_{ng})t} e^{-i(\omega_{B_j n}+\omega_{ng})\tau'} e^{-i\omega_{ng}\tau} \\
	&= e^{i\omega_{B_j}t} e^{-i\omega_{B_j}\tau'} e^{-i\omega_{ng}\tau},
\end{align*}
where we used $\omega_{B_j n} + \omega_{ng} = \omega_{B_j g} = \omega_{B_j}$, having set $E_g = 0$ as the energy reference. The field phases (Contraction A, signal at $t_2$, idler at $t_1$) become
\begin{align*}
	e^{-i\omega_s t_2} e^{-i\omega_i t_1} &= e^{-i(\omega_s+\omega_i)t} e^{i(\omega_s+\omega_i)\tau'} e^{i\omega_i\tau}.
\end{align*}

Collecting by variable, the $t$-dependent overall phase $e^{i(\omega_{B_j} - \omega_s - \omega_i)t}$ factors out, and the remaining integrand separates. In the new variables, Eq.~\eqref{eq:A3} becomes
\begin{equation}
	\label{eq:A4}
	\begin{split}
		\mathcal{M}_j^{(A)}(t) = &-\frac{1}{\hbar^2} \sum_n \iint d\omega_s\, d\omega_i\; \Psi(\omega_s, \sigma_2;\, \omega_i, \sigma_1) \\
		&\times\bra{B_j}\hat{d}_{\sigma_2}\ket{n}\bra{n}\hat{d}_{\sigma_1}\ket{g}\; e^{i(\omega_{B_j} - \omega_s - \omega_i)t} \\
		&\times\int_0^\infty\! d\tau'\; e^{[i(\omega_s+\omega_i-\omega_{B_j})-\gamma_B]\tau'} \\
		&\times\int_0^\infty\! d\tau\; e^{[-i(\omega_{ng}-\omega_i)-\gamma_n]\tau}.
	\end{split}
\end{equation}

Both integrals are over positive time differences with exponential damping, so both converge automatically ($\gamma_n, \gamma_B > 0$).

Final-state integral ($\tau'$):
\begin{equation}
	\label{eq:A5}
	\begin{split}
		&\int_0^\infty d\tau'\,
		e^{[i(\omega_s+\omega_i-\omega_{B_j})-\gamma_B]\tau'} \\
		&\qquad= \frac{-i}{(\omega_{B_j}-\omega_s-\omega_i)-i\gamma_B}
		\equiv L(\omega_s+\omega_i).
	\end{split}
\end{equation}
This is the final-state Lorentzian enforcing two-photon energy conservation.

Intermediate-state integral ($\tau$):
\begin{equation}
	\label{eq:A6}
	\int_0^\infty d\tau\,
	e^{[-i(\omega_{ng}-\omega_i)-\gamma_n]\tau}
	= \frac{-i}{(\omega_{ng}-\omega_i)-i\gamma_n}.
\end{equation}
which is the retarded intermediate-state propagator. In Contraction A the photon absorbed at $t_1$ has frequency $\omega_i$, so the intermediate-state propagator is evaluated at $\omega_i$.

Combining all factors, Contraction A gives:
\begin{subequations}
	\label{eq:A7}
	\begin{equation}
		\label{eq:A7a}
		\begin{split}
			\mathcal{M}_j^{(A)} &\propto \iint d\omega_s d\omega_i \Psi(\omega_s, \sigma_2; \omega_i, \sigma_1) \\
			&\quad\times\sum_n \frac{\bra{B_j}\hat{d}_{\sigma_2}\ket{n} \bra{n}\hat{d}_{\sigma_1}\ket{g}} {[\hbar(\omega_{ng}{-}\omega_i){-}i\hbar\gamma_n]} L(\omega_s {+} \omega_i).
		\end{split}
	\end{equation}

	For Contraction B (idler at $t_2$, signal at $t_1$), the JSA component is $\Psi(\omega_s, \sigma_1; \omega_i, \sigma_2)$ [Eq.~\eqref{eq:11}], indices swapped and the field phases are $e^{-i\omega_i t_2} e^{-i\omega_s t_1}$. The only change is in the $\tau$-integral, where the photon absorbed at $t_1$ now has frequency $\omega_s$ instead of $\omega_i$: \[ \int_0^\infty d\tau\, e^{[-i(\omega_{ng}-\omega_s)-\gamma_n]\tau} = \frac{-i}{(\omega_{ng}-\omega_s)-i\gamma_n}. \]

	The $\tau'$-integral is unchanged (it depends only on the total photon energy $\omega_s + \omega_i$). Thus,
	\begin{equation}
		\label{eq:A7b}
		\begin{split}
			\mathcal{M}_j^{(B)} &\propto \iint d\omega_s d\omega_i \Psi(\omega_s, \sigma_1; \omega_i, \sigma_2) \\
			&\quad\times\sum_n \frac{\bra{B_j}\hat{d}_{\sigma_2}\ket{n} \bra{n}\hat{d}_{\sigma_1}\ket{g}} {[\hbar(\omega_{ng}{-}\omega_s){-}i\hbar\gamma_n]} L(\omega_s {+} \omega_i).
		\end{split}
	\end{equation}
\end{subequations}
Adding both contractions, the total amplitude $\mathcal{M}_j = \mathcal{M}_j^{(A)} + \mathcal{M}_j^{(B)}$ shares the common final-state Lorentzian $L(\omega_s+\omega_i)$. Defining the frequency-dependent two-photon matrix element $T_j(\omega)$ [Eq.~\eqref{eq:12}] absorbs the $\hbar$ prefactors and the sum over intermediate states, yielding the total amplitude [Eq.~\eqref{eq:13}].

The nondegenerate excitation scheme places the signal at the exciton resonance ($\omega_s = E_X/\hbar$) and the idler off-resonance by the biexciton binding energy ($\omega_i = \omega_s - E_B/\hbar$). The frequency-dependent kernel $T(\omega)$ is a Lorentzian centered on $\omega_X$:
\begin{equation}
	\label{eq:A8}
	T(\omega) = \frac{d_0 d_{XX,X}}{\hbar(\omega_X - \omega) - i\hbar\gamma_X},
\end{equation}
where $d_0 = \bra{X_K}\hat{d}_+\ket{g}$ is the ground-to-exciton transition dipole matrix element (Eq.~\eqref{eq:8}) and $d_{XX,X} = \bra{B_2}\hat{d}_-\ket{X_K}$ is the exciton-to-biexciton transition dipole. Equation~\eqref{eq:19} gives the suppression ratio for each binding energy and yields $\approx 0.006$ for $\Phi_1$.

A signal-photon detuning changes the value of this Lorentzian but not the phase prefactors. When $|\omega_X-\omega_s|\gg\gamma_X$, the pair-absorption rate decreases as $\gamma_X^2/(\omega_X-\omega_s)^2$ [Eq.~\eqref{eq:12}]. We set $\omega_s=\omega_X$ to maximize the rate.

\section{\label{sec:appB}Frequency Integral for Type-II SPDC}

With $T_2(\omega_s) = T_3(\omega_s) \equiv T(\omega_s)$ under valley symmetry, the dominant-contraction amplitude from the main text (Sec.~\ref{sec:dominant}) is
\begin{equation*}
	\mathcal{M}_2 \approx \frac{1}{\sqrt{2}}\iint d\omega_s d\omega_i f(\omega_s,\omega_i) T(\omega_s) L(\omega_s {+} \omega_i).
\end{equation*}
For a CW (monochromatic) pump at $\omega_p = \omega_s + \omega_i$, which is valid when the pump linewidth is much narrower than the biexciton linewidth $\gamma_{B}$ (a condition readily satisfied by the single-mode CW lasers used to pump SPDC sources), the JSA takes the standard form~\cite{gu2020}:
\begin{equation}
	\label{eq:B1}
	f(\omega_s, \omega_i) = \mathcal{N} \delta(\omega_s {+} \omega_i {-} \omega_p) \mathrm{sinc}[(\omega_X {-} \omega_s)T_e/2],
\end{equation}
where $\mathrm{sinc}[(\omega_X {-} \omega_s)T_e/2]$ is the marginal signal spectral amplitude and $\mathcal{N}$ is the source amplitude set by the finite pump envelope and photon flux. Substitution into the amplitude:
\begin{align*}
	\mathcal{M}_2 &= \frac{\mathcal{N}}{\sqrt{2}}\iint d\omega_s d\omega_i \delta(\omega_s {+} \omega_i {-} \omega_p) \\
	&\quad\times\mathrm{sinc}[(\omega_X {-} \omega_s)T_e/2]\, T(\omega_s) L(\omega_s {+} \omega_i).
\end{align*}
The $\delta$-function eliminates the $\omega_i$ integration, setting $\omega_i = \omega_p - \omega_s$ and hence $\omega_s + \omega_i = \omega_p$. The final-state factor becomes a constant $L(\omega_p)$:
\begin{align*}
	\mathcal{M}_2 &= \frac{\mathcal{N} L(\omega_p)}{\sqrt{2}} \int_{-\infty}^{\infty} d\omega_s \mathrm{sinc}[(\omega_X {-} \omega_s)T_e/2]\, T(\omega_s).
\end{align*}
We change the integration variable from $\omega_s$ to the signal detuning from the intermediate exciton, $\Delta \equiv \omega_X - \omega_s$:
\begin{align*}
	\mathcal{M}_2 &= \frac{\mathcal{N} L(\omega_p)}{\sqrt{2}} \int_{-\infty}^{\infty} d\Delta \mathrm{sinc}(\Delta T_e/2)\, T(\omega_X {-} \Delta).
\end{align*}
The valley selection rules reduce the intermediate-state sum in $T(\omega)$ [Eq.~\eqref{eq:12}] to a single exciton $\ket{X_K}$ at frequency $\omega_X$, giving
\begin{equation*}
	T(\omega_X-\Delta)
	= \frac{d_0d_{XX,X}}{\hbar(\Delta-i\gamma_X)}.
\end{equation*}
Substituting:
\begin{equation}
	\label{eq:B2}
	\mathcal{M}_2 = \frac{\mathcal{N} L(\omega_p) d_0 d_{XX,X}} {\sqrt{2}\hbar} \int_{-\infty}^{\infty} d\Delta \frac{\mathrm{sinc}(\Delta T_e/2)}{\Delta - i\gamma_X}.
\end{equation}
The source-dependent prefactor $\mathcal{N} L(\omega_p)/\sqrt{2}$ is common to all pathways and is absorbed into the rate prefactor $\kappa_\alpha$.

For the physical sinc spectral amplitude, we represent the sinc as a Fourier integral over a rectangular temporal window: \[ \mathrm{sinc}\left(\frac{\Delta T_e}{2}\right) = \frac{1}{T_e}\int_{-T_e/2}^{T_e/2} e^{i\Delta t} dt. \] Substituting into the spectral integral Eq.~\eqref{eq:B2} and exchanging the order of integration, the $\Delta$-integral evaluates by contour integration. The sole pole at $\Delta = i\gamma_X$ lies in the upper half-plane, while the convergence factor $e^{i\Delta t}$ decays in the upper (lower) half-plane for $t > 0$ ($t < 0$). Closing in the upper half-plane for $t > 0$ picks up the pole; closing in the lower half-plane for $t < 0$ encloses no poles:
\begin{align*}
	\int_{-\infty}^{\infty} \frac{e^{i\Delta t}}{\Delta - i\gamma_X} d\Delta = \begin{cases} 2\pi i e^{-\gamma_X t}, & t > 0, \\
		0, & t < 0. \end{cases}
\end{align*}
Only the $t > 0$ portion of the rectangular window $[-T_e/2, T_e/2]$ contributes, leaving an elementary integral: \[ \int_{0}^{T_e/2} e^{-\gamma_X t} dt = \frac{1}{\gamma_X}\left(1 - e^{-\gamma_X T_e/2}\right). \] Collecting the prefactor $2\pi i/T_e$ from the sinc representation, the spectral integral evaluates to
\begin{equation}
	\label{eq:B3}
	\int_{-\infty}^{\infty} d\Delta \frac{\mathrm{sinc}(\Delta T_e/2)}{\Delta - i\gamma_X} = \frac{2\pi i}{T_e \gamma_X} \left(1 - e^{-\gamma_X T_e/2}\right).
\end{equation}
Substituting Eq.~\eqref{eq:B3} into Eq.~\eqref{eq:B2} gives the full amplitude $\mathcal{M}_2$, from which we identify the effective matrix element by grouping the source-dependent factors $\mathcal{N} L(\omega_p) (2\pi/T_e)$ into the prefactor that becomes $\kappa_\alpha$:
\begin{equation*}
	\mathcal{M}_2 = \frac{2\pi\,\mathcal{N} L(\omega_p)}{\sqrt{2}\,T_e} \frac{d_0 d_{XX,X}}{\hbar} \frac{i}{\gamma_X} \left(1 - e^{-\gamma_X T_e/2}\right).
\end{equation*}
The resonant two-photon amplitude is therefore
\begin{equation}
	\label{eq:B4}
	\mathcal{T} = \frac{d_0 d_{XX,X}}{\hbar} \frac{i}{\gamma_X}\left(1 - e^{-\gamma_X T_e/2}\right),
\end{equation}
which is Eq.~\eqref{eq:26} of the main text.

The $\ket{B_3}$ pathway proceeds through $\ket{X_{K'}}$ instead of $\ket{X_K}$, with $\sigma^+ \leftrightarrow \sigma^-$ polarizations. By time-reversal symmetry at zero magnetic field, all dipole matrix elements and dephasing rates are identical under $K \leftrightarrow K'$, yielding
\begin{equation}
	\label{eq:B5}
	\mathcal{T}^{(\sigma^-\sigma^+)} = \mathcal{T}^{(\sigma^+\sigma^-)} = \mathcal{T}.
\end{equation}

\section{\label{sec:appC}Eigenstate-Resolved Pumping Rate}

We now derive the resonance-weighted pumping-rate formula Eq.~\eqref{eq:27} by projecting the Dyson-series amplitudes onto the eigenstate basis.

The Dyson-series amplitudes derived in Appendices~\ref{sec:appA} and~\ref{sec:appB} give the transition amplitude from the ground state to each biexciton configuration. For a pure biphoton polarization state $\ket{\psi_\mathrm{2ph}}_\mathrm{pol} = \sum_{\mu_1\mu_2} c_{\mu_1\mu_2}\ket{\mu_1\mu_2}$, each polarization pair selects the configuration $B_{\mu_1\mu_2}$. Here, $(\mu_1,\mu_2)$ labels either a cross-circular or a cocircular polarization pair defined after Eq.~\eqref{eq:27}. Each amplitude is proportional to $c_{\mu_1\mu_2}\mathcal{T}$ [Eq.~\eqref{eq:B5}]. The configuration-basis superposition imprinted by the photon pair is therefore
\begin{equation}
	\label{eq:C1}
	\ket{\psi_\mathrm{prep}} \propto \sum_{\mu_1\mu_2} c_{\mu_1\mu_2} \ket{B_{\mu_1\mu_2}},
\end{equation}
where $\mathcal{T}$ and all source-dependent spectral factors have been suppressed temporarily.

The transition amplitude from the ground state to eigenstate $\ket{\Phi_\alpha}$, denoted $\widetilde{\mathcal{M}}_\alpha$, is obtained by projecting the configuration-basis state onto $\bra{\Phi_\alpha} = \sum_j \braket{\Phi_\alpha | B_j}\bra{B_j}$ (Table~\ref{tab:eigenstates}) and restoring the suppressed spectral prefactors. In doing so, the final-state Lorentzian $L(\omega_s + \omega_i)$ from the configuration-basis derivation [Eq.~\eqref{eq:A5}] must be evaluated at the eigenstate energy $\omega_{\Phi_\alpha g}$ rather than the configuration energy $\omega_{B_j g}$, because the true energy eigenstates of the biexciton manifold are the exchange-split states $\ket{\Phi_\alpha}$. This gives
\begin{equation}
	\label{eq:C2}
	\widetilde{\mathcal{M}}_\alpha = \frac{2\pi\,\mathcal{N}}{\sqrt{2}\,T_e} L_\alpha(\omega_p)\mathcal{T} \sum_{\mu_1\mu_2} c_{\mu_1\mu_2} \braket{ \Phi_\alpha | B_{\mu_1\mu_2} },
\end{equation}
with $L_\alpha(\omega_p) = [(\omega_{\Phi_\alpha g}-\omega_p)-i\gamma_{B}]^{-1}$. The transition probability per incident pair is therefore
\begin{widetext}
	\begin{equation}
		\label{eq:C3}
		\begin{split}
			P_\alpha = |\widetilde{\mathcal{M}}_\alpha|^2 &= \frac{2\pi^2\,|\mathcal{N}|^2}{T_e^2} |L_\alpha(\omega_p)|^2 |\mathcal{T}|^2 \left| \sum_{\mu_1\mu_2} c_{\mu_1\mu_2} \braket{ \Phi_\alpha | B_{\mu_1\mu_2} } \right|^2 \\
			&= \frac{2\pi^2\,|\mathcal{N}|^2}{T_e^2} |L_\alpha(\omega_p)|^2 |\mathcal{T}|^2 \sum_{\mu_1\mu_2,\mu_1'\mu_2'} \braket{ \Phi_\alpha | B_{\mu_1\mu_2} } \braket{ B_{\mu_1'\mu_2'} | \Phi_\alpha } c_{\mu_1\mu_2} c_{\mu_1'\mu_2'}^*.
		\end{split}
	\end{equation}

	For a mixed biphoton state $\rho = \sum_k p_k \ket{\psi_k}\bra{\psi_k}$, the ensemble average replaces $c_{\mu_1\mu_2}c_{\mu_1'\mu_2'}^*$ by the polarization density matrix $\rho_{\mu_1\mu_2,\mu_1'\mu_2'}$. Using the linear-in-flux scaling of ETPA in the isolated-pair regime~\cite{raymer2021,schlawin2013nc}, the pumping rate is
	\begin{equation}
		\label{eq:C4}
		W_\alpha = \kappa_\alpha \sum_{\mu_1\mu_2,\mu_1'\mu_2'} \braket{ \Phi_\alpha | B_{\mu_1\mu_2} } \braket{ B_{\mu_1'\mu_2'} | \Phi_\alpha } \rho_{\mu_1\mu_2,\mu_1'\mu_2'},
	\end{equation}
	with \[ \kappa_\alpha \equiv \kappa_0\, \frac{\gamma_{B}^2} {(\omega_{\Phi_\alpha g}-\omega_p)^2+\gamma_{B}^2}, \qquad \kappa_0 \equiv R_\mathrm{pairs} \frac{2\pi^2\,|\mathcal{N}|^2 |\mathcal{T}|^2}{T_e^2\,\gamma_{B}^2}. \]
\end{widetext}
This form separates the eigenstate geometry from the light correlations, as before. It also shows that the final-state resonance depends on the biexciton eigenstate.

\section{\label{sec:appD}Coherence Preservation and Visibility Estimates}

\subsection{\label{sec:appD1}Tripartite state, decoherence factor, and mean dwell time}

After the signal photon is absorbed from $\ket{\psi(\varphi)}_\mathrm{Bell}$ [Eq.~\eqref{eq:30}], but before the idler arrives, the joint state of (material + idler photon + environment) is

\begin{equation}
	\label{eq:D1}
	\begin{split}
		\ket{\psi_\mathrm{tot}(t)} &= \frac{1}{\sqrt{2}}\Big( \ket{X_K}\ket{\sigma^-}_i\ket{\mathcal{E}_K(t)} \\
		&\quad + e^{i\varphi}\ket{X_{K'}}\ket{\sigma^+}_i \ket{\mathcal{E}_{K'}(t)}\Big),
	\end{split}
\end{equation}

where $\ket{\mathcal{E}_K(t)}$ and $\ket{\mathcal{E}_{K'}(t)}$ are the environmental states conditioned on the exciton valley, evolving from a common initial state $\ket{\mathcal{E}_0}$ at the moment of absorption ($t = 0$). The coherence between the two branches is encoded in the off-diagonal element of the density matrix obtained by tracing over the environment:

\begin{equation}
	\label{eq:D2}
	\begin{split}
		\rho_{K,K'}^{(\mathrm{mat+ph})}(t) &= \frac{e^{-i\varphi}}{2} \braket{\mathcal{E}_K(t) | \mathcal{E}_{K'}(t) },
	\end{split}
\end{equation}

where $\rho_{K,K'}^{(\mathrm{mat+ph})}$ denotes the coherence in the joint (material + idler photon) reduced state. The factor $\braket{ \mathcal{E}_K(t) | \mathcal{E}_{K'}(t) }$ quantifies how distinguishable the environment has become at time $t$.

To model the time dependence of $\braket{ \mathcal{E}_K(t) | \mathcal{E}_{K'}(t) }$, we treat the environment as a bath that acquires which-valley information through acoustic-phonon-mediated dephasing, and adopt a Markovian exponential model in which the environmental overlap satisfies

\begin{equation}
	\label{eq:D3}
	\frac{d}{dt}\braket{ \mathcal{E}_K(t) | \mathcal{E}_{K'}(t) } = -\gamma_\mathrm{valley}\braket{ \mathcal{E}_K(t) | \mathcal{E}_{K'}(t) },
\end{equation}

where $\gamma_\mathrm{valley}$ is the effective rate at which the environment distinguishes the two valleys. Integrating Eq.~\eqref{eq:D3} gives
\begin{equation}
	\label{eq:D4}
	\braket{ \mathcal{E}_K(t) | \mathcal{E}_{K'}(t) } = e^{-\gamma_\mathrm{valley} t},
\end{equation}
the standard decoherence-function form.

Before averaging the decoherence factor, we need a model for the distribution of the intermediate-state delay. We construct this model from two ingredients in Appendix~\ref{sec:appB}. First, the Fourier representation of the sinc joint spectral amplitude limits $\tau$ to the causal interval $[0,\, T_e/2]$. Second, the retarded propagator for the intermediate exciton [Eq.~\eqref{eq:A2}] supplies an exponential factor $e^{-\gamma_X\tau}$.

As a modeling estimate, we adopt a normalized truncated-exponential distribution on this interval,
\begin{equation}
	\label{eq:D5}
	p(\tau) = \frac{e^{-\gamma_X\tau}} {\displaystyle\int_0^{T_e/2} e^{-\gamma_X\tau'}\, d\tau'} = \frac{\gamma_X\, e^{-\gamma_X\tau}} {1 - e^{-\gamma_X T_e/2}}, \quad \tau \in [0,\, T_e/2].
\end{equation}

Using this $p(\tau)$, we average the decoherence factor. Within the valley-symmetric limit, the two pathways share the same phase-independent material amplitude $\mathcal{T}$ [Eqs.~\eqref{eq:25} and~\eqref{eq:26}]. The joint biexciton--environment state for a fixed delay $\tau$ is
\begin{equation}
	\label{eq:D6}
	\ket{\Psi(\tau)} = \frac{\mathcal{T}}{\sqrt{2}}\left(\ket{B_2}\ket{\mathcal{E}_K(\tau)} + e^{i\varphi}\ket{B_3}\ket{\mathcal{E}_{K'}(\tau)}\right),
\end{equation}
where $\ket{\mathcal{E}_K(\tau)}$ and $\ket{\mathcal{E}_{K'}(\tau)}$ are the environment states after dwelling for time $\tau$ with the intermediate exciton in valley $K$ and $K'$, respectively. The biexciton reduced density matrix $\rho_\mathrm{BX}(\tau) \equiv \mathrm{Tr}_\mathcal{E}[\ket{\Psi(\tau)}\bra{\Psi(\tau)}]$ then has configuration-basis matrix elements
\begin{subequations}
	\label{eq:D7}
	\begin{align}
		\braket{B_2|\rho_\mathrm{BX}(\tau)|B_2} &= \braket{B_3|\rho_\mathrm{BX}(\tau)|B_3} = \tfrac{|\mathcal{T}|^2}{2}, \label{eq:D7a} \\
		\begin{split}
			\braket{B_2|\rho_\mathrm{BX}(\tau)|B_3} &= \tfrac{|\mathcal{T}|^2}{2}\,e^{-i\varphi}\,\braket{\mathcal{E}_K(\tau)|\mathcal{E}_{K'}(\tau)} \\&= \tfrac{|\mathcal{T}|^2}{2}\,e^{-i\varphi}\,e^{-\gamma_\mathrm{valley}\tau},
		\end{split} \label{eq:D7b}
	\end{align}
\end{subequations}
where the final equality uses Eq.~\eqref{eq:D4}. The diagonal populations are $\tau$-independent, while the off-diagonal coherence carries the decoherence factor $e^{-\gamma_\mathrm{valley}\tau}$ acquired during the dwell.

Averaging over $\tau$ with $p(\tau)$ [Eq.~\eqref{eq:D5}] therefore leaves the populations unchanged and suppresses the off-diagonal coherence by the dwell-time-averaged overlap, which we denote with an overline:

\begin{equation}
	\label{eq:D8}
	\begin{split}
		\overline{\braket{ \mathcal{E}_K | \mathcal{E}_{K'} }}
		&\equiv \int_0^{T_e/2} p(\tau)\, \braket{\mathcal{E}_K(\tau) | \mathcal{E}_{K'}(\tau)}\, d\tau \\
		&= \int_0^{T_e/2} p(\tau)\, e^{-\gamma_\mathrm{valley}\tau}\, d\tau \\
		&= \frac{\gamma_X}{\gamma_X {+} \gamma_\mathrm{valley}} \frac{1 - e^{-(\gamma_X + \gamma_\mathrm{valley})T_e/2}} {1 - e^{-\gamma_X T_e/2}}.
	\end{split}
\end{equation}

Because the diagonal populations carry no $\varphi$ dependence while the off-diagonal coherence carries $e^{-i\varphi}\,\overline{\braket{\mathcal{E}_K|\mathcal{E}_{K'}}}$, the model gives $W_\mathrm{bright}(\varphi) \propto 1 + \overline{\braket{\mathcal{E}_K|\mathcal{E}_{K'}}}\cos\varphi$. The dwell-time-averaged overlap is therefore the visibility, $V_\mathrm{deph} = \overline{\braket{\mathcal{E}_K|\mathcal{E}_{K'}}}$, giving Eq.~\eqref{eq:47}.

The same distribution gives the mean intermediate-state dwell time:
\begin{equation}
	\label{eq:D9}
	\langle\tau\rangle
	\equiv \int_0^{T_e/2}\tau\,p(\tau)\,d\tau
	= \frac{1}{\gamma_X}
	- \frac{T_e/2}{e^{\gamma_XT_e/2}-1}.
\end{equation}
which is Eq.~\eqref{eq:46}. In the narrow-bandwidth limit ($\gamma_X T_e \gg 1$), the second term vanishes and $\langle\tau\rangle \to 1/\gamma_X$. In the broad-bandwidth limit ($\gamma_X T_e \ll 1$), expanding the exponential to second order gives $\langle\tau\rangle \to T_e/4$. This mean characterizes the adopted distribution and enters the Sec.~\ref{sec:decoherence} estimates through the averaged overlap $V_\mathrm{deph}$ [Eqs.~\eqref{eq:47} and~\eqref{eq:D8}].

\subsection{\label{sec:appD2}Combined visibility from intervalley scattering}

Unlike valley dephasing, which degrades the off-diagonal coherence without changing the pathway population, intervalley scattering physically transfers the intermediate exciton from one valley to the other at rate $\gamma_\mathrm{iv}$. An exciton created in valley $K$ at dwell time $\tau = 0$ remains in $K$ with probability $e^{-\gamma_\mathrm{iv}\tau}$. Averaging this factor over the same dwell-time distribution $p(\tau)$ [Eq.~\eqref{eq:D5}] used for the valley-dephasing calculation gives
\begin{equation}
	\label{eq:D10}
	\begin{split}
		\overline{e^{-\gamma_\mathrm{iv}\tau}} &= \int_0^{T_e/2} p(\tau)\,e^{-\gamma_\mathrm{iv}\tau}\, d\tau \\
		&= \frac{\gamma_X}{\gamma_X + \gamma_\mathrm{iv}} \frac{1 - e^{-(\gamma_X + \gamma_\mathrm{iv})T_e/2}} {1 - e^{-\gamma_X T_e/2}},
	\end{split}
\end{equation}
which is identical in form to Eq.~\eqref{eq:D8} with the replacement $\gamma_\mathrm{valley} \to \gamma_\mathrm{iv}$. The fraction of each pathway that is scattered into the wrong valley is therefore
\begin{equation}
	\label{eq:D11}
	P_\mathrm{iv} = 1 - \overline{e^{-\gamma_\mathrm{iv}\tau}} = 1 - \frac{\gamma_X}{\gamma_X + \gamma_\mathrm{iv}} \frac{1 - e^{-(\gamma_X + \gamma_\mathrm{iv})T_e/2}} {1 - e^{-\gamma_X T_e/2}},
\end{equation}
which is Eq.~\eqref{eq:48}.

The scattered fraction converts each pathway into the wrong valley:
\begin{subequations}
	\label{eq:D12}
	\begin{align}
		\ket{X_K} &\to \sqrt{1 - P_\mathrm{iv}}\ket{X_K} + \sqrt{P_\mathrm{iv}} e^{i\chi_K}\ket{X_{K'}}, \label{eq:D12a} \\
		\ket{X_{K'}} &\to \sqrt{1 - P_\mathrm{iv}}\ket{X_{K'}} + \sqrt{P_\mathrm{iv}} e^{i\chi_{K'}}\ket{X_K}, \label{eq:D12b}
	\end{align}
\end{subequations}
where $\chi_K$, $\chi_{K'}$ are random phases associated with the scattering event. After the second photon is absorbed, the $(\sigma^+,\sigma^-)$ pathway produces $\ket{B_2}$ with probability $1-P_\mathrm{iv}$ and the corrupted cocircular configuration $\ket{B_6}$ with probability $P_\mathrm{iv}$; similarly the $(\sigma^-,\sigma^+)$ pathway produces $\ket{B_3}$ or the corrupted configuration $\ket{B_1}$. The corrupted configurations do not interfere with the cross-circular sector $\{\ket{B_2},\ket{B_3}\}$, so they add a phase-independent background without affecting the cross-circular coherence.

\section{\label{sec:appE}Estimates of the Residual Bright Population}

At zero field, time-reversal symmetry gives $T_2(\omega)=T_3(\omega)$. A stray magnetic field produces the valley splitting $\Delta E_Z=g\mu_B B$ with $|g|\approx4$~\cite{srivastava2015,aivazian2015}. A field of $B=1$~mT gives $\Delta E_Z\approx0.2~\mu$eV. Expanding the Lorentzian kernel [Eq.~\eqref{eq:A8}] around resonance yields
\begin{equation*}
	\left|\frac{\delta T}{T}\right|\approx\frac{\Delta E_Z}{\hbar\gamma_X}\approx10^{-4},
\end{equation*}
which gives a residual bright population $|\delta T/T|^2\approx10^{-8}$. Magnetic disorder and pump-induced valley polarization depend on the sample and excitation conditions. We therefore use $|\delta T/T|=10^{-2}$ as a conservative model value. This value contributes $10^{-4}$ to $P_\mathrm{res}$.

The wave-plate estimates use the conservative tolerances $\theta_\mathrm{err}=1^\circ$ and $\delta\approx0.06$~rad ($\lambda/100$), including chromatic variation across the $17$~nm pair bandwidth. These values give the contributions listed in Table~\ref{tab:budget}.

To obtain the visibility associated with $P_\mathrm{res}$, we normalize the rates at $\varphi=0$ and $\varphi=\pi$ to their sum. The normalized rates are then $W_\mathrm{bright}(0)\propto1-P_\mathrm{res}$ and $W_\mathrm{bright}(\pi)\propto P_\mathrm{res}$. Their contribution to the visibility is
\begin{equation*}
	V_\mathrm{res}\equiv
	\frac{W_\mathrm{bright}(0)-W_\mathrm{bright}(\pi)}
	{W_\mathrm{bright}(0)+W_\mathrm{bright}(\pi)}
	=1-2P_\mathrm{res}.
\end{equation*}
This gives the factor $V_\mathrm{res}$ used in Eq.~\eqref{eq:56}.

\section{\label{sec:appF}Signal Estimate, Backgrounds, and Readout Noise Floors}

\subsection{Coincidence-transmission readout}

Let $P_\mathrm{1ph}\approx10^{-1}$ be the one-photon absorption probability at the $A$-exciton resonance~\cite{li2014optical}. For $R_\mathrm{pairs}=10^{10}~\mathrm{s}^{-1}$, the single-exciton creation rate is $R_X\approx10^{9}~\mathrm{s}^{-1}$. The estimated uncorrelated sequential rate is $R_X^2\tau_X\approx2\times10^{6}~\mathrm{s}^{-1}$ for $\tau_X\sim2$~ps~\cite{robert2016}. Although the one-photon and sequential backgrounds can be large, they are independent of $\varphi$. Their mean contributions are removed by comparing the measurements at $\varphi=0$ and $\varphi=\pi$.

Exciton--exciton interactions and elastic scattering also give $\varphi$-independent backgrounds. Their mean values are removed by the phase comparison.

We estimate the phase-dependent coincidence loss that can be resolved from the Poisson uncertainty of the true and accidental coincidence counts. A measurement of duration $T_\mathrm{int}=1$~s at either $\varphi=0$ or $\varphi=\pi$ collects
\begin{equation}
	\label{eq:F1}
	N_c=R_\mathrm{pairs}\eta_\mathrm{det}^2T_\mathrm{int}\approx10^{8}
\end{equation}
true transmitted coincidences. The accidental count is $N_a=N_c/\mathrm{CAR}$. Assuming a coincidence window $\tau_w\approx100$~ps, $\mathrm{CAR}=1/(R_\mathrm{pairs}\tau_w)\approx1$ and hence $N_a\approx N_c$.

Pair absorption changes the true coincidence count between $\varphi=0$ and $\varphi=\pi$ by $\Delta P_\mathrm{abs}N_c$. The accidental counts do not carry this phase-dependent signal, but they contribute to the shot noise. Each phase measurement has Poisson variance $N_c+N_a$ in the weak-absorption limit. The two measurements are independent, so the variance of their difference is $2(N_c+N_a)$. The signal-to-noise ratio is therefore
\begin{equation*}
	\mathrm{SNR}_\mathrm{coinc}
	=\frac{\Delta P_\mathrm{abs}N_c}{\sqrt{2(N_c+N_a)}},
\end{equation*}
which gives Eq.~\eqref{eq:57}. With $N_a\approx N_c$, the fractional shot-noise floor relative to the true coincidences is
\begin{equation}
	\label{eq:F2}
	\sigma_\mathrm{dip}
	=\frac{\sqrt{2(N_c+N_a)}}{N_c}
	\approx\frac{2}{\sqrt{N_c}}
	\approx2\times10^{-4}.
\end{equation}
With the pump on $\Phi_3$, $\kappa_\mathrm{bright}\approx0.43\,\kappa_3$. The difference between the two dip depths is therefore $\Delta P_\mathrm{abs}=0.57\,P_\mathrm{abs}(\pi)$. Requiring this difference to exceed $\sigma_\mathrm{dip}$ gives
\begin{equation}
	\label{eq:F3}
	P_\mathrm{abs}(\pi)\gtrsim3.5\times10^{-4}.
\end{equation}
This threshold corresponds to $R_\mathrm{pairs}P_\mathrm{abs}(\pi)\approx3.5\times10^6~\mathrm{s}^{-1}$ and is comparable to the sequential rate estimated above. Since $N_c\propto T_\mathrm{int}$, the noise floor decreases as $1/\sqrt{T_\mathrm{int}}$.

\subsection{Pump-probe readout}

The dominant pump-probe background is the probe-transmission change produced by single excitons from one-photon absorption. Its mean is removed by comparing the probe transmission changes measured at $\varphi=0$ and $\varphi=\pi$. At each phase, $\Delta T/T$ is obtained from the two measurements $T_\mathrm{on}$ and $T_\mathrm{off}$. The comparison uses $T_\mathrm{on}(0)$, $T_\mathrm{off}(0)$, $T_\mathrm{on}(\pi)$, and $T_\mathrm{off}(\pi)$. Each measurement counts $N=R_\mathrm{probe}T_\mathrm{int}$ probe photons, where $R_\mathrm{probe}$ is the detected probe-photon rate.

For a weak probe response, the photon counts in $T_\mathrm{on}$ and $T_\mathrm{off}$ are both approximately $N$. Their difference has Poisson variance $2N$, so the variance of $\Delta T/T$ is $2/N$. The two phase measurements are independent, giving a variance $4/N$. The probe shot-noise floor is therefore
\begin{align}
	\label{eq:F4}
	\sigma_\mathrm{pp}
	&=\frac{2}{\sqrt{R_\mathrm{probe}T_\mathrm{int}}}\approx6.3\times10^{-8},
\end{align}
for $R_\mathrm{probe}\sim10^{15}~\mathrm{s}^{-1}$ and one second of integration. This gives Eq.~\eqref{eq:58}.

\bibliography{references}

\end{document}